\documentclass[12pt]{spieman}  

\usepackage{amsmath,amsfonts,amssymb}
\usepackage{graphicx}
\usepackage{setspace}
\usepackage{tocloft}
\usepackage{subfig}
\usepackage{amsmath}

\usepackage{lipsum}
\usepackage{float}
\usepackage{amsmath}
\usepackage{latexsym}
\usepackage{booktabs,etoolbox}
\usepackage[utf8x]{inputenc}
\usepackage{caption}
\usepackage{xspace}

\usepackage{lineno}

\let\oldAA\AA
\renewcommand{\AA}{\text{\normalfont\oldAA}}

\title{Emu: A Case Study for TDI-like Imaging for Infrared Observation from Space}

\author[a]{Joice Mathew}
\author[a]{James Gilbert}
\author[a]{Robert Sharp}
\author[a]{Alexey Grigoriev}
\author[b]{Adam D. Rains}
\author[b,c]{Anna M. Moore}
\author[a]{Annino Vaccarella}
\author[a,d]{Aurelie Magniez}
\author[a]{David Chandler}
\author[a]{Ian Price}
\author[b]{Luca Casagrande}
\author[b,e,f]{Maru{\v s}a {\v Zerjal}}
\author[b]{Michael Ireland}
\author[b]{Michael S. Bessell}
\author[a]{Nicholas Herrald}
\author[a]{Shanae King}
\author[b,g]{Thomas Nordlander}

\affil[a]{Advanced Instrumentation and Technology Centre, Research School of Astronomy and Astrophysics, Australian National University, Canberra, ACT 2611, Australia}
\affil[b] {Research School of Astronomy and Astrophysics, Australian National University, Canberra, ACT 2611, Australia}
\affil[c] {Australian National University Institute for Space, Australian National University, Canberra, ACT, Australia}
\affil[d] {University of Rennes 1, Rue du Thabor, 35000 Rennes, France}
\affil[e] {Instituto de Astrof{\'{\i}}sica de Canarias, E-38205 La Laguna, Tenerife, Spain}
\affil[f] {Universidad de La Laguna, Dpto. Astrof{\'{\i}}sica, E-38206 La Laguna, Tenerife, Spain}
\affil[g] {ARC Centre of Excellence for All Sky Astrophysics in 3 Dimensions (ASTRO 3D), Canberra, ACT 2611, Australia}

\cftpagenumbersoff{figure}
\cftpagenumbersoff{table}

\begin{document} 
\maketitle

\begin{abstract}

A wide-field zenith-looking telescope operating in a mode similar to Time-Delay-Integration (TDI) or drift scan imaging can perform an infrared sky survey without active pointing control but it requires a high-speed, low-noise infrared detector. Operating from a hosted payload platform on the International Space Station (ISS), the \textit{Emu} space telescope employs the paradigm-changing properties of the Leonardo SAPHIRA electron avalanche photodiode array to provide powerful new observations of cool stars at the critical water absorption wavelength (1.4 $\mu$m) largely inaccessible to ground-based telescopes due to the Earth’s own atmosphere. 
Cool stars, especially those of spectral-type M, are important probes across contemporary astrophysics, from the formation history of the Galaxy to the formation of rocky exoplanets. Main sequence M-dwarf stars are the most abundant stars in the Galaxy and evolved M-giant stars are some of the most distant stars that can be individually observed. The Emu sky survey will deliver critical stellar properties of these cool stars by inferring oxygen abundances via measurement of the water absorption band strength at 1.4 $\mu$m. 
Here we present the TDI-like imaging capability of Emu mission, its science objectives, instrument details and simulation results.

\end{abstract}


{\noindent \footnotesize\textbf{*}Corresponding author,  \linkable{joice.mathew@anu.edu.au}}

\begin{spacing}{1}   


\section{Introduction}
\label{sect:intro}  

Even though M-dwarfs are the most abundant stars in our Galaxy\cite{M-dwarf_most_abundant_star}, relatively little is known about their make-up (i.e., their abundance patterns of elements heavier than Hydrogen and Helium) because of their intrinsic faintness, cool temperatures ($T_{\rm eff}$ $\lesssim$  4,000 K), and complex atmospheres. Their low photospheric temperatures result in a stellar spectrum that contains overlapping atomic and molecular absorption, making meaningful measurement of bulk stellar metallicity challenging---let alone key elemental abundance patterns. This complexity makes for a highly degenerate problem, rendering it difficult to uniquely determine a set of stellar parameters to describe a given star, something routinely done for Solar-type stars. In such cool stars properties like temperature ($T_{\rm eff}$), the bulk metallicity ([M/H]), and elemental abundances like [O/H] and [C/H] all prominently affect the large-scale shape of the observed spectrum and greatly complicate the use of traditional analysis techniques. 

Most of the dominant molecular absorbers in cool atmospheres contain Oxygen (e.g. TiO in the optical; H$_2$O and CO in the near-infrared) where, in chemical equilibrium conditions, the amount of Oxygen-free to form such molecules is sensitive to the Carbon abundance through the formation of energetically favourable CO \cite{burrows_sharp_1999, Veyette+2016}. The observational consequence of this is that both Oxygen and Carbon---elements difficult to measure on their own---have an out-sized impact on the shape of a spectrum and greatly affect the ability to reliably measure stellar metallicity \cite{Veyette+2016}. 

The ability to reliably measure Oxygen would thus aid in resolving this degeneracy and help to unlock the chemistry of the most common kind of star. To this end, one particular wavelength region of interest is the H$_2$O band at 1.4 $\mu$m. Unfortunately, this region is exceedingly challenging to observe from the ground due to moisture in our atmosphere\cite{Mauna_Kea_water_absorption_NIR, IRTF}. However, space-based observations show tremendous promise in opening this unique window into the physics of stellar atmospheres of cool stars.

The International Space Station (ISS) provides a unique platform to perform life, physical, Earth, and space science research and technology development. The ISS, being a science platform, provides all of the resources that a scientific instrument would expect from an Earth-like laboratory such as power, cooling, data, and communication needs. The ISS external payload platform provides a new opportunity to realize space missions extraordinarily fast and of low cost in comparison to traditional research conducted on-board the ISS. It provides the flexibility to be replaced with different instruments or, in some cases, brought back to the Earth. The experiments on these platforms can fit into short-term funding cycles available on national and multi-national levels. External payload accommodations are available on the ``Expedite the Processing of Experiments to the Space Station" (ExPRESS) Logistics Carrier (ELC), ``Japanese Experiment Module-Exposed Facility" (JEM-EF) and the``Columbus-External Payload Facility" (Columbus-EPF) of the ISS\cite{ISS_external_payload_guide}. The commercially available external platforms for small payloads include NanoRacks External Platform \cite{nanoracks} (operational from 2016) attached to JEM and Bartolomeo platform \cite{Bartelomeo_DLR_payload} attached to Columbus Module, which is operated by Airbus Defence and Space. Emu leverages these emerging cost-effective and reduced mission risk platforms to perform the infrared sky survey, along with key detector technology demonstration in space.

The Emu mission will undertake its near-infrared sky survey in a band centred on 1.4 $\mu$m (Emu $W$-band or $W_{\rm E}$) as well as in $J$ band (for cross-calibration) as a method of estimating oxygen abundance in the atmospheres of cool stars down to a magnitude m$_H$(AB) $\le$ 12.5. Emu is a compact wide-field photometer destined for a 6-month mission on the exterior of the ISS. The Emu payload conforms to a 6U CubeSat form factor and employs ``noise-free" SAPHIRA infrared detector array\cite{Jamie_SAPHIRA}, and will perform Time-Delay-Integration (TDI) like imaging from the ISS\cite{Emu_SPIE_Proc}. The ISS presents a novel platform for a survey instrument, due to its precessing orbit. It follows that a high-speed imager with a $1.2^{\circ}$ diagonal field of view (FOV) toward the zenith can map the sky without active pointing. Therefore, it demands that stars do not traverse more than one pixel per readout, to avoid smearing. These frames are then shifted, aligned, and stacked to produce a continuous strip with an effective exposure time equivalent to the full transit time of the array. Emu will also  demonstrate the near-infrared TDI-like imaging capability of SAPHIRA in space, which can play a crucial role in future missions such as  GaiaNIR\cite{Gaia_white_paper, GAIA-NIR_1, GAIA-NIR_2}. 
Emu is in the early phase of development and the conceptual design of the instrument has been completed (at the time of writing). The launch has not been finalised yet and we are exploring different opportunities to fly this mission. Critical hardware and subsystems for the mission are being developed and validated.

\subsection{TDI-like imaging}

TDI or ‘drift scan’ imaging is an electronic scan technique, based on the concept of the accumulation of cumulative exposures of the same object as it is moving across the field of view\cite{TDI_basics}. One of the major telescope facilities that use the TDI imaging technique is the Sloan Digital Sky Survey (SDSS) telescope located at Apache Point Observatory\cite{SDSS}. While the CMOS readout architecture of the Leonardo eAPD precludes conventional TDI imaging operations, the low effective readout noise of the eAPD (when operated with avalanche gain) means that quasi-TDI observations can be implemented effectively by reading the full array at the pixel crossing time and shifting and stacking images (without avalanche gain, conventional CMOS readout noise levels would result in a highly read noise limited final observation, with little to no sensitivity).
The major advantage of TDI mode operations is the avoidance of the need for a beam steering and stabilization tracking mirror. Conventional CMOS array operations would require the incoming field to be stabilised onto the focal plane array for the full integration time. This requires a complex (assumed open-loop) control system, and also introduces additional demands on image stability and sensitivity variation across the tracked field. The use of a tracking mirror system would, however, likely relax requirements on full-field static distortion and reduce the intermediate data rate. TDI will require considerable parallel onboard data processing to reduce the value-added frame data volume.

\section{Scientific Background} \label{Scientific Background}

The metal content of a star, often referred to as its `metallicity' or [M/H], is used to describe its chemical composition and refers to the relative amount of material other than Hydrogen and Helium. Chemical composition---both in terms of the bulk metallicity and individual elemental abundances---is one of a star's fundamental properties (alongside mass, age, and angular momentum), and the measurement of which has been a critical component of stellar and Galactic astrophysics for many decades now. There are now many hundreds of thousands of stars with measured metallicities and elemental abundances in massive spectroscopic surveys like GALAH\cite{de_Silva+2015} and APOGEE\cite{Majewski+2017} but, despite their prevalence, most of these aren't M-dwarfs. Indeed, GALAH has a temperature cut around $\sim$ 4,000 K below which it doesn't produce stellar parameters as spectral models become increasingly unreliable and the determination of a unique set of stellar parameters increasingly challenging.

In the field of stellar astrophysics, the Iron abundance [Fe/H] is commonly used as a proxy for the stellar bulk metallicity. Iron can be reliably measured in Solar-type stars, and it typically serves as the entry point for investigations into stellar chemistry or galactic chemical evolution, especially when paired with other elemental abundances or stellar properties like age. However, Iron can be difficult to measure accurately in the atmospheres of cooler stars like M-dwarfs \cite{Rojas-Ayala, Bonfils+05, Bean+06, Terrien+2012, Newton+2014, Mann+2015, Rains+21}, where metallicity-abundance degeneracies paired with model limitations pose substantial barriers. In M-giant stars, the problem
becomes worse because of non-Local Thermodynamic Equilibrium (non-LTE) effects that become so severe for
electronic transitions that even the stellar temperature is difficult to measure \cite{Ireland}. Oxygen makes up a higher proportion of the “metal” composition of stars than iron and so is in principle more
representative of the metallicity of stars. However, its atomic abundance is much harder to measure \cite{Asplund}. It has few spectral lines; the permitted high excitation OI triplet (0.7771 - 0.7774\,$\mu$m) is affected by non-LTE
and inhomogeneities in solar-type and cooler stars, while the extremely weak forbidden lines [OI] at 0.6300\,$\mu$m  and
0.6363\,$\mu$m require high sensitivity and high-resolution spectroscopy to separate them from the forest of transitions
associated with other elements and is not practical in M stars.

\begin{figure}[h]
\begin{center}
\includegraphics[scale=0.33]{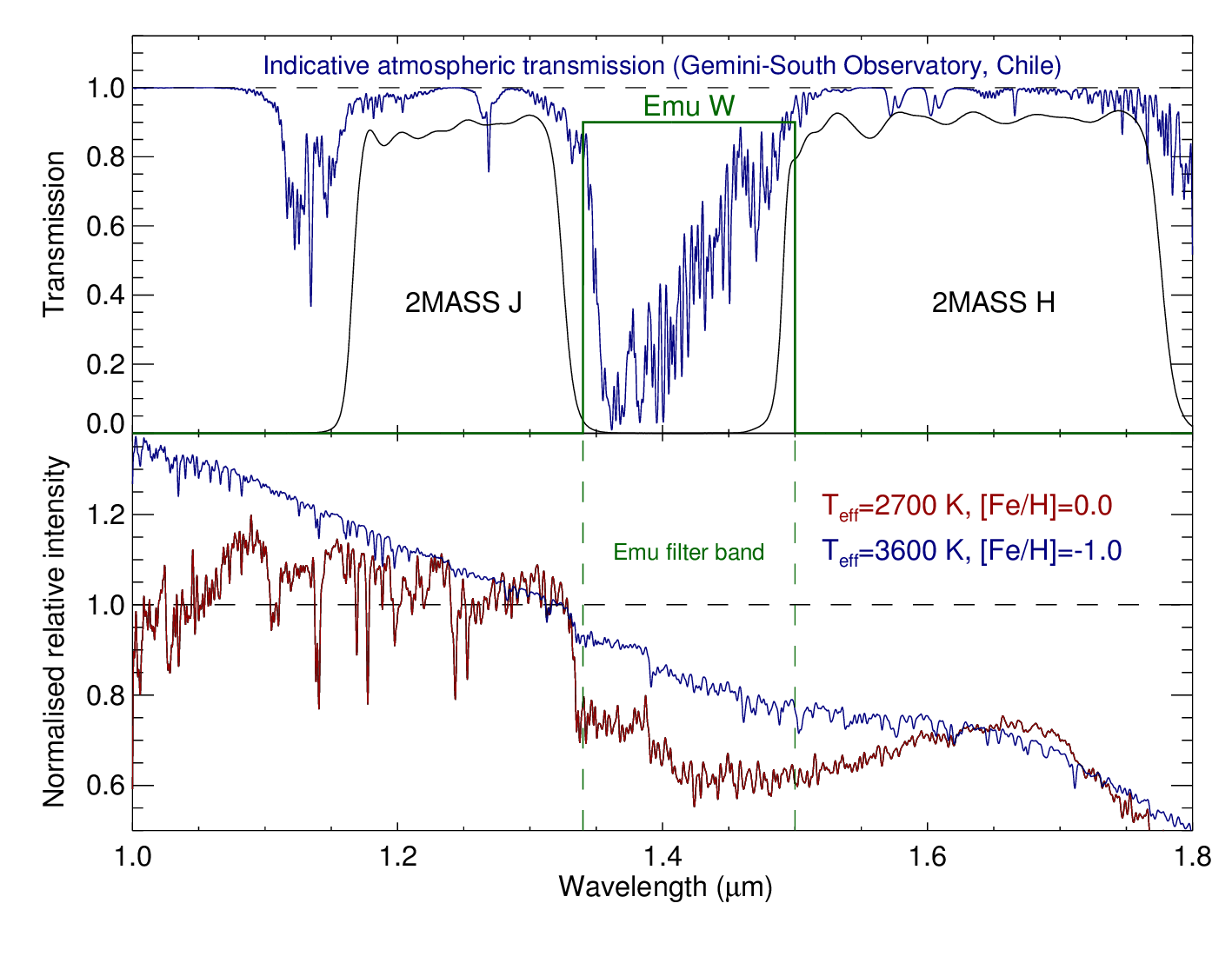}
\end{center}
\caption{\textbf{Top:} Earth's atmospheric transmission compared to the 2MASS $J$ and $H$ bands\cite{Skrutskie+06}, as well as the proposed Emu $W$-band. Note that it is not possible to effectively observe the region between $J$ and $H$ from the ground due to water absorption (Green box in the top panel), thus demonstrating the strength of Emu. \textbf{Bottom:} MARCS model \cite{Gustafsson} synthetic spectra for two M dwarfs in the same wavelength region, one with $T_{\rm eff}=2700\,$K and [Fe/H] = 0.0, and the other with T$_{\rm eff} =3600\,$K and [Fe/H] = -1. Note how the targeted wavelengths demonstrate different [Fe/H] sensitivities as compared to the adjacent $J$ and $H$ bands, useful in constraining stellar parameters.}
\label{fig:Emu 'Water band'}
\end{figure}

\begin{figure}[h]
\begin{center}
\includegraphics[scale=0.99]{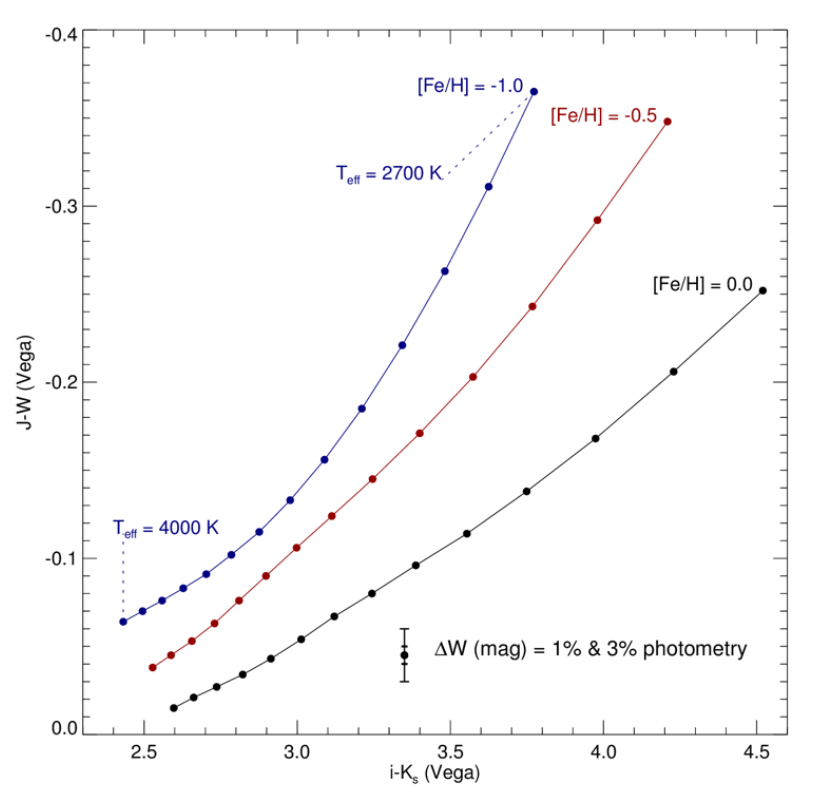}
\end{center}

\caption{A synthetic color-color diagram shows the ability of the Emu $W_{\rm E}$-band photometry (the data product from
our all-sky survey) to measure stellar metallicity. Tracks show the zero-age stellar main sequence for cool-stars (2700
$\leq$  $T_{\rm eff}$ $\leq$ 4000 K, in steps of 100 K) at the indicated [Fe/H] metallicity with all elemental ratios such as [O/Fe] fixed to Solar values. Note how the use of a $J-W$ colour enables the separation of the stars into distinct metallicity sequences, demonstrating the power of the Emu $W_{\rm E}$-band. The $i-Ks$ colour on the X-axis is a proxy for stellar temperature (with some dependence on [Fe/H]), with the $J-W$ colour plotted on the Y-axis being primarily sensitive to [Fe/H].}
\label{fig:color color diagram}
\end{figure}

The strongest water absorption feature in the coolest ($<$ 3000 K) stars is at 2.7 $\mu$m. Observation at this wavelength is challenging due to the high thermal background. The strongest absorption feature that can be readily accessed is between the astronomical $J$ and $H$-band windows at 1.4 $\mu$m. Its
measurement from the ground is challenging due to the dominance, and rapid variability, of the water
absorption column in the Earth’s atmosphere and the wide field of view required to provide sufficient reference
stars for a differential calibration. Space-based observations are therefore desired. The $W_{\rm E}$-band centre is at 1.4 $\mu$m with a bandwidth of 0.15 $\mu$m. The band is strongest in the atmospheres of the coolest stars ($T_{\rm eff}$ = 2700 K is adopted as a functional limit based on currently available model atmosphere data) and weakest in warmer stars (with $T_{\rm eff}$ = 4000 K adopted as the functional upper limits for cool stars). The band strength is also dependent on the relative abundance of heavy elements in the star. Emu will measure the water abundance by comparing the observed flux of stars in the Emu $W_{\rm E}$-filter to the expected flux predicted based on broad-band photometric values available in the literature.

The Emu $W_{\rm E}$-band provides an improved indicator
of the underlying oxygen abundance than other broad-band filters (which blends complex metallicity signatures) as the dominant molecular absorber at 1.4 $\mu$m is H$_2$O. Compare this to the optical, where there is a larger contribution from a much greater range of sources: from atoms, as well as molecules such as oxides (e.g. TiO, VO, ZrO), and hydrides (e.g. MgH, CaH, SiH), which all contribute to the complex metallicity and abundance signatures.
Two example synthetic spectra are shown in Fig.~\ref{fig:Emu 'Water band'}, showing the variation in the water absorption strength for two
temperature and metal abundance levels. The
synthetic spectrum of a warm (shown in blue, effective temperature, $T_{\rm eff}$ = 3600 K) low-metallicity M-star is shown
compared to a cooler star with higher metallicity (shown in red). The Emu $W$-filter band, indicated by the green lines,
straddles the strong water absorption band in both cool stars, but with significantly more absorption in the higher
metallicity source. O is fully condensed into H$_2$O in dwarf stars at solar metallicity 3000 K or in giant stars at 2000 K\cite{Tsuji}. For near dwarfs with $T_{\rm eff}$ below 3300 K, the depth of water features are therefore more of a measure of [O/H] than $T_{\rm eff}$. A representative Earth's atmospheric transmission spectrum (for ground-based astronomy\footnote{Infrared spectra of the atmospheric transmission above Mauna Kea are shown here- https://www.gemini.edu/observing/telescopes-and-sites/sites}) is also shown,
as are the canonical and widely used $J$ and $H$ band astronomical filters\cite{Mauna_Kea_filter_set}. The water absorption in the Earth’s atmosphere hinders these
observations from being performed with ground-based telescopes\cite{IRTF}. A full statistical analysis, using all available photometric data will of
course underpin the measurement of stellar abundance using the $W_{\rm E}$-band. However, a simple color-color diagram
(photometric intensity ratios in different filter bands) demonstrates the power of our unique Emu $W_{\rm E}$-band for
determining stellar atmospheric abundance (Fig.~\ref{fig:color color diagram}) across a wider range of stellar parameters. A star's
location in the color-color space made available by the $W_{\rm E}$-band provides a probe of the atmospheric
water content, and hence oxygen abundance, free from the degeneracies associated with other broad-band
filters alone. Photometric accuracy at the 3\% level (signal-to-noise ratio $\geqslant$ 30) will provide powerful empirical evidence for the metal abundance of the coolest stars. For warmer atmospheres and higher metallicities, the brighter sources (with 1\% photometric errors) from the Emu survey will be used. Emu will provide critical empirical evidence of the validity of the state-of-the-art models, which have significant
uncertainty at high metallicities and for the warmer stars.

\section{Instrument Overview} \label{Instrument Overview}

\begin{figure}[h!]
\centering
\includegraphics[scale=0.55]{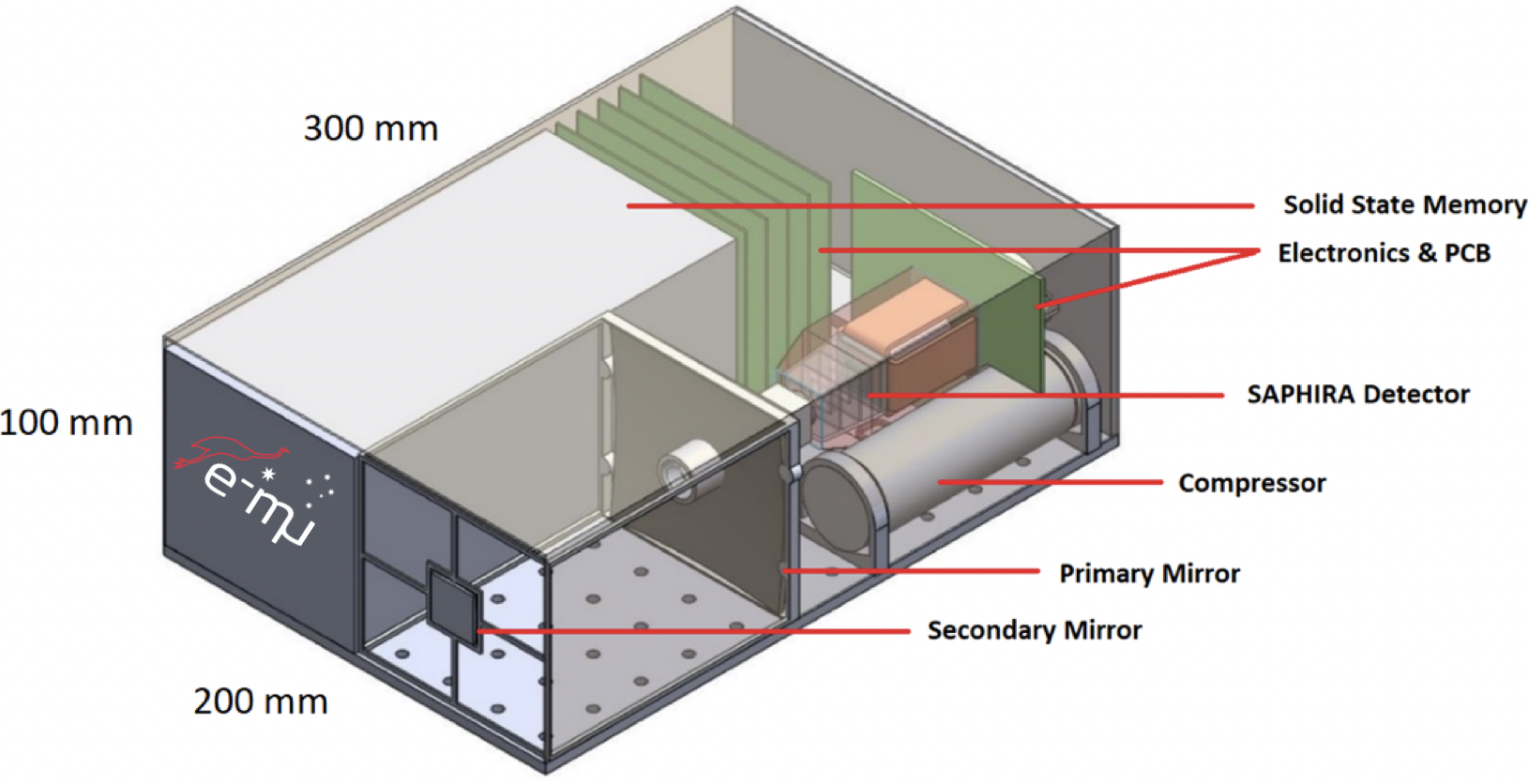}
\caption{The Emu conceptual model is shown. The optical elements, including the cold detector assembly and thermal management system, occupy a volume equivalent to 3U. A further 1U is required for the control system electronics. In the baseline design, a further 2U is set aside for a data mass storage system.}
\label{fig:Emu conceptual model}
\end{figure}

Emu is a 6U form factor payload to be hosted on the ISS. The instrument layout of Emu is shown in Fig. \ref{fig:Emu conceptual model}.

\begin{table}[h]
\caption{The Emu instrument specifications are shown here. 
The optimum pixel size for the Emu sky survey is $10^{\prime\prime}$. A Full-Width Half Maximum (FWHM) of 2 pixels is considered to satisfy the Nyquist sampling and this will result in a spatial resolution of $20^{\prime\prime}$ for Emu. Emu uses a low readout noise SAPHIRA detector to enable TDI-like imaging capability. SAPHIRA will be operated in a Correlated Double Sampler (CDS) readout mode resulting in a science frame rate of 25 Hz. In this mode, a Non-Destructive Readout (NDR) reset and exposure frame will be taken, which will result in a 50 Hz NDR frame rate. Emu will fit in a 6U volume and it can achieve a sensitivity around m$_H$(AB)  $\sim$ 12.5 in H band with a signal to noise ratio of 10.}
\begin{center}
\begin{tabular}{ll}
\hline 
\rule[-1ex]{0pt}{3.5ex} Instrument & NIR imager \\
\rule[-1ex]{0pt}{3.5ex} Imaging mode & TDI-like  \\
\rule[-1ex]{0pt}{3.5ex} Field of view & $0.71^{\circ}$ $\times$ $0.89^{\circ}$\\
\rule[-1ex]{0pt}{3.5ex} Aperture size &  83 mm $\times$ 83 mm  \\
\rule[-1ex]{0pt}{3.5ex} Central obstruction & $\sim$ 12.5 $\%$  \\
\rule[-1ex]{0pt}{3.5ex} Equivalent circular aperture diameter &  93.65 mm \\
\rule[-1ex]{0pt}{3.5ex} Effective area &  $\sim$ 60 cm$^2$ \\
\rule[-1ex]{0pt}{3.5ex} Focal length & 495 mm  \\
\rule[-1ex]{0pt}{3.5ex} Operating bands & $W_{\rm E}$ (1.34 - 1.5 $\mu$m) and $J$ (1.17 - 1.33 $\mu$m) \\
\rule[-1ex]{0pt}{3.5ex} Detector & SAPHIRA eAPD from Leonardo MW Ltd.  \\
\rule[-1ex]{0pt}{3.5ex} Array size & 320 x 256 \\
\rule[-1ex]{0pt}{3.5ex} Spatial resolution & FWHM = $20^{\prime\prime}$ (2 pixels)\\
\rule[-1ex]{0pt}{3.5ex} Frame rate & 25 Hz CDS, (50 Hz NDR) \\
\rule[-1ex]{0pt}{3.5ex} Number of field transits & 256 \\
\rule[-1ex]{0pt}{3.5ex} Effective exposure time & $\sim$ 7.2 sec (in $W_{\rm E}$-band) and  $\sim$ 2.4 sec (in $J$-band)\\
\rule[-1ex]{0pt}{3.5ex} Limiting magnitude (single orbit) & m$_H$ (AB) $\le$ 12.5 in $H$-band (with SNR =10)\\
\rule[-1ex]{0pt}{3.5ex} Weight & $\le$ 8 kg \\
\rule[-1ex]{0pt}{3.5ex} Dimension (L $\times$ W$\times$ H)& $300 \times 200\times100$ mm (6U form factor) \\
\rule[-1ex]{0pt}{3.5ex} Power & $ < 50$  W \\
\hline
\end{tabular}
\label{table:instrument details}
\end{center}
\end{table}

The optical system is a compact Cassegrain telescope, with a re-imaged pupil and it will occupy 2U space. The integrated detector cooler assembly  (IDCA)  system contains the cold baffle, detector, and Stirling cooler. The IDCA will fit in a 1U unit. The IDCA holds the filters which are required to achieve the desired passband and as well as thermal requirements. A further 0.5U is required for the ANU ``Rosella" detector readout electronics system. Rosella is a modular and compact readout electronics system to read out the SAPHIRA detector. In the baseline design, a further 2U is set aside for a commercially available off-the-shelf (COTS) data mass storage system. The different instrument parameters are shown in Table. \ref{table:instrument details}. The drifts scan survey approach,
enabled by the low readout noise of the SAPHIRA detector at a high frame rate, removes the mass and complexity
associated with a conventional pointing-and-tracking telescope. The telescope's primary mirror will be baffled and
sufficient thermal blocking is achieved in the cold section of the instrument. The SAPHIRA detector will be cooled to
80 K (in principle SAPHIRA can be operated at a higher temperature, but this will cause an increase in the dark current\cite{SAPHIRA_radiation_testing}), and while this aspect of the project presents a significant design challenge, the power consumption and
waste heat dissipation necessary are within the scope of capabilities provided by our preferred host platform on the ISS.

The direction of flight will be along the short axis of the detector (256 pixels) and the transverse field will be along the long axis (320 pixels). Along with the $W_{\rm E}$-band imaging, Emu will also undertake a sky survey in the \textit{J}-band for cross-calibration of data. The simultaneous $W_{\rm E}$ and \textit{J}-band imaging is achieved by placing two separate $W_{\rm E}$ and \textit{J} filters before the focal plane. The filters will be placed such that they will be perpendicular to the direction of flight, which would allow scanning the same portion of the sky in both bands. $J$ and $W_{\rm E}$-band will occupy 60 and 180 pixels with an intermediate dead band of 16 pixels necessary for mounting the filters at the focal plane ($\sim 0.4$ mm gap between filters). The mission will operate for 6 months on the ISS with the majority of data recorded with on-board storage for sample return, due to communication bandwidth restrictions. We will be able to downlink $\sim$ 1 \% of this data for calibration purposes.

\section{ISS Platform and Jitter} \label{ISS platform}

The ISS is maintained in a nearly circular orbit with an average altitude of 400 km, at an inclination of $\sim$ 51.6 degrees to Earth's equator. The orbital period of ISS is $\sim$ 92.68 minutes, with $\sim$ 15.54 orbits per day\cite{ISS_external_payload_guide}. Emu will perform the near-infrared sky survey between $+51.6^{\circ}$ and $-51.6^{\circ}$ declination during its six months of operation and thus would scan $\sim$78 \% of the total sky.

\begin{figure}[h]
\begin{center}
\includegraphics[scale=0.62]{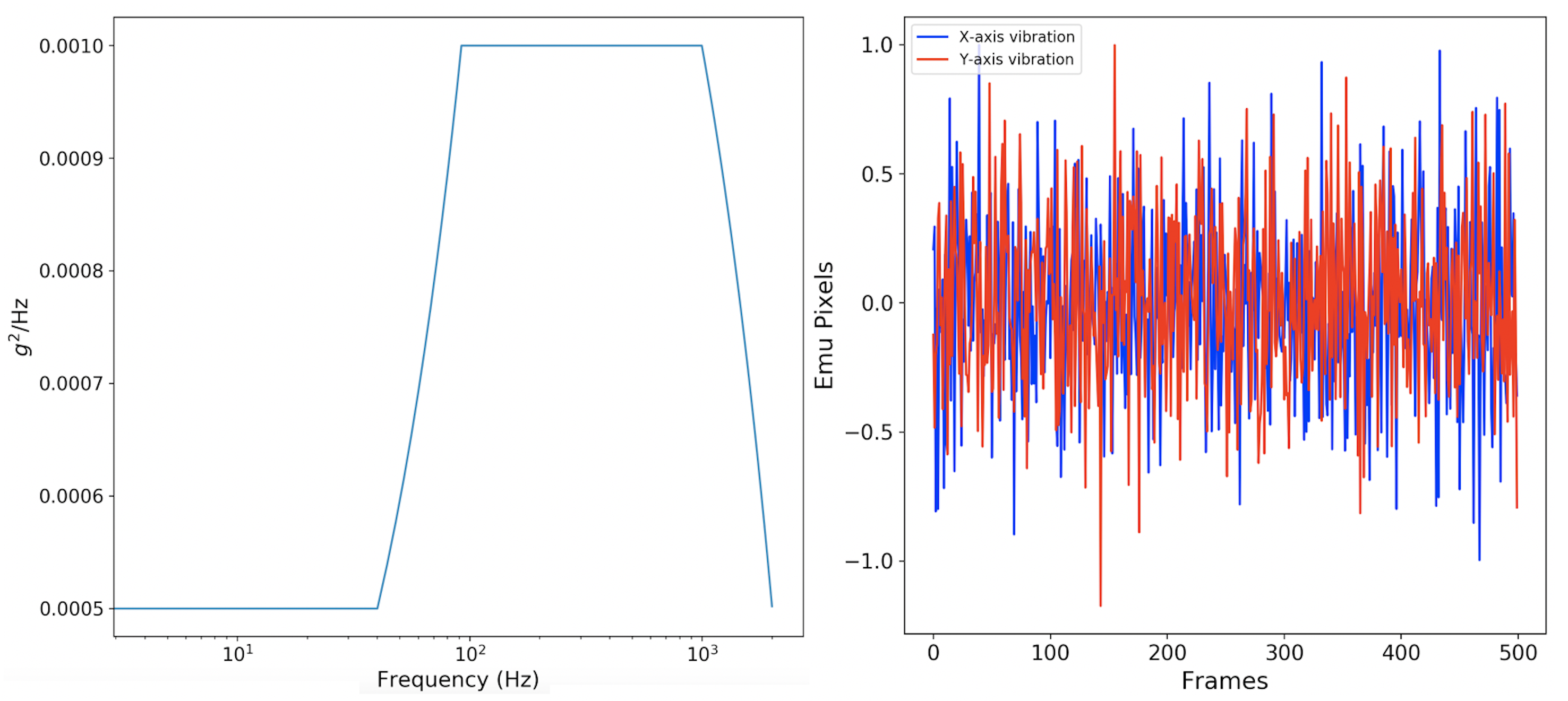}

\end{center}
\caption{{\it Left}: ISS platform PSD. \, {\it Right}: Random vibrations in Emu pixel scale. Note the ISS jitter is less than one Emu pixel scale and thus it will not be a critical noise source in the Emu data output.}

\label{fig:PSD_and_vibration}
\end{figure}

In order to understand the effect on image quality of Emu, due to ISS jitter, we have simulated a random vibration profile from the power spectral density (PSD) data available from the External Payloads Proposer’s Guide to the ISS\cite{ISS_external_payload_guide}. This corresponds to around 30 micro radians (RMS) in both axes of the focal plane and less than one Emu pixel ($10^{\prime\prime} $). The ISS platform PSD and the random vibrations (in Emu pixel scale) derived from the PSD are shown in Fig.\ref{fig:PSD_and_vibration}. One of the missions which performed the characterization of the ISS vibration environment is the Optical Payload for Lasercomm Science (OPALS)\cite{OPALS}. OPALS is a technology demonstration for laser communication aboard the ISS and it has been utilized for different extended missions after its primary mission. From this experiment, the ISS jitter has been estimated to be in 20-30 micro radians range \cite{ISS_jitter_OPALS}. This is in agreement with the derived jitter from the ISS power spectral density (PSD) data.

Thus the ISS jitter will not be a critical noise source in the Emu data and just be within the tolerance (FWHM= 20"). Also, the jitter does not introduce significant rotational issues in the data; but does introduce a significant global translation between frames. We can freeze out the vibration frame to frame (because of fast frame readout), but they do have an impact on the image trajectory across multiple frames of the data. Cross-correlation between successive frames will be performed to shift and stack multiple frames (details in Section \ref{Simulation}).

\section{Instrument Performance Estimates} \label{Simulation}

Traditionally, TDI uses a CCD technology by shifting the electrical charge across the detector at the same velocity as the object image to avoid blurring. The analog of this for CMOS technology such as SAPHIRA is taking images at a high enough frame-rate, and post-stacking them to produce effectively long exposures. The conventional CMOS can't work in TDI mode, because of the high readout noise at a fast frame rate. However, the low electronic noise of the SAPHIRA detector allows the camera to read out images at the same rate that stars transit the image pixel (25 Hz). A series of images is recorded as the sky crosses the static (with respect to the Space Station) telescope field of view. The images are then aligned (using integer pixel shifts) in post-processing and stacked to give a composite image. The exposure time of each frame is chosen to match the time taken for a star to traverse one pixel on the detector, such that the next exposure will see the same image but with everything shifted by one whole column. The frame rate is set by the drift speed, which is $\sim$$233 ^{\prime\prime}$ s$^{-1}$ for the ISS in a 400 km LEO.

\begin{table}[h]
\caption{Simulation input parameters}
\begin{center}
\begin{tabular}{ll}
\hline 
\rule[-1ex]{0pt}{3.5ex} Telescope effective collecting area & 55.1 cm$^2$ \\
\rule[-1ex]{0pt}{3.5ex} Plate scale & $\sim 416.7 ^{\prime\prime} $ \,mm$^{-1}$\\
\rule[-1ex]{0pt}{3.5ex} Optics efficiency & $\sim0.8$\\
\rule[-1ex]{0pt}{3.5ex} Wavelength bands &  $W_{\rm E}$ and $J$ \\
\rule[-1ex]{0pt}{3.5ex} Number of pixels in the direction of flight & 256 pixels \\
\rule[-1ex]{0pt}{3.5ex} Number of pixels in the transverse direction & 320 pixels \\
\rule[-1ex]{0pt}{3.5ex} Pixel size &  24 $\mu$m  \\
\rule[-1ex]{0pt}{3.5ex} Avalanche gain & 45  \\
\rule[-1ex]{0pt}{3.5ex} Effective readout noise & 0.978 $e^-$  \\
\rule[-1ex]{0pt}{3.5ex} Dark current & 8.4 $e^{-}$\,s$^{-1}$ \\

\rule[-1ex]{0pt}{3.5ex} Quantum efficiency & 0.7 in $W_{\rm E}$-band and 0.6 in $J$-band\cite{ESO_readnoise} \\
\rule[-1ex]{0pt}{3.5ex} Thermal background rate & 1 $ e^-$\,s$^{-1}$\\

\rule[-1ex]{0pt}{3.5ex} Single frame exposure time  & 0.04 s \\
\rule[-1ex]{0pt}{3.5ex} Angular speed of the platform & $\sim$$233 ^{\prime\prime}$ s$^{-1}$\\

\hline
\end{tabular}
\label{table:Simulation input parameters}
\end{center}
\end{table}

The exposure time per individual integration is then calculated from the pixel size, $S_{Pix}$, (in units of arc seconds) and drift speed, $R_{p}$. 

\begin{equation}
\tau_{exp} = \frac{S_{Pix}}{R_{p}}
\end{equation}

The frame rate is given by

\begin{equation}
\mathrm{Frame}\,\mathrm{rate}\  = \frac{1}{\tau_{exp}}
\label{eq:frame rate}
\end{equation}

The frame rate for a $10^{\prime\prime}$ pixel based on Equation. \ref{eq:frame rate} is 23.3 Hz. For simplicity, for the Emu baseline design we have considered 25 Hz frame rate.

The full effective exposure time is then a function of the number of pixels along the flight path of the platform, $N_{x}$

\begin{equation}
\tau_{eff} = N_x \,\tau_{exp}
\end{equation}


A simulation has been performed to recreate the effects of the SAPHIRA detector, ROSELLA electronics, optical elements, background noises, ISS vibrations, TDI-like imaging, and stars. The different input parameters to simulate Emu mission data are shown in Table. \ref{table:Simulation input parameters}. The simulation input parameters provide high flexibility and would allow simulating similar TDI-like mission (for example, GaiaNIR mission - see Section \ref{GaiaNIR mission concept} for more details).

The thermal background at the focal plane is another input parameter which is discussed in section \ref{Thermal emission analysis}. Orbital and spacecraft platform parameters for the simulation consist of the angular speed of the ISS and the data on vibrations on the ISS external platform (discussed in section \ref{ISS platform}). To simulate the star-field we have used the 2MASS star catalog data\cite{2MASS}.

\begin{figure}[h]
\begin{center}
\includegraphics[scale=.55]{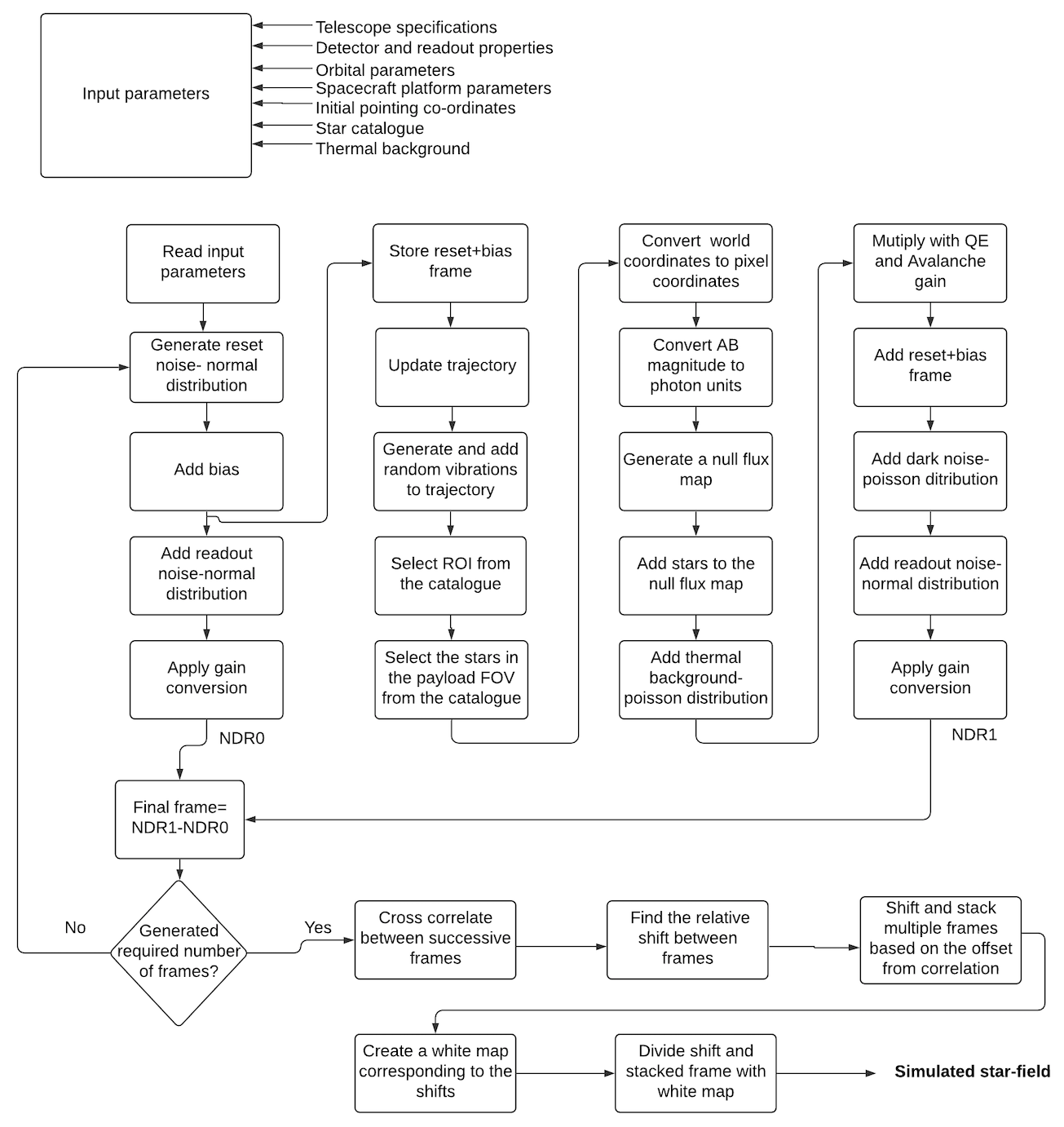}
\end{center}
\caption{Flow chart of Emu simulation}
\label{fig:Flow chart of Emu simulation}
\end{figure}

The different processing steps and logical flow of
the simulator is schematically shown as a flowchart in Fig. \ref{fig:Flow chart of Emu simulation} (see section \ref{Appendix}). The SAPHIRA operates on correlated double sample (CDS) readout mode over the full array with a resulting science framerate of $\sim$ 25 Hz.  In this process Non-Destructive Readout Zero (NDR0) frame or reset frame will be generated followed by an exposed frame or Non-Destructive Readout One (NDR1) frame. Each of these frames requires two readouts (one after reset and one after exposure) that are differenced to produce a CDS science image.

In order to simulate the vibrations on the ISS external platform, a random vibration profile is generated as mentioned in Section \ref{ISS platform} and it is then added to the trajectory. Based on the resulted trajectory coordinates, a region of interest (ROI) is selected from the 2MASS star catalog\cite{Skrutskie+06}. From this ROI, the stellar coordinates and magnitude data are extracted for stars that would fall in the Emu FOV. In order to add these selected stars to the frame, a coordinate transformation is applied to convert the world coordinates to pixel coordinates. 

In this simulation 2MASS, $J$, and $H$ band magnitudes are used to simulate the flux of stars. The 2MASS $J$ and $H$-band Vega-based magnitudes are converted to the AB system using the Equation. \ref{eq:J band to AB} and \ref{eq:H band to AB}

\begin{equation} \label{eq:J band to AB} 
M_{AB}= J-0.91
\end{equation}

\begin{equation} \label{eq:H band to AB} 
M_{AB}= H-1.39
\end{equation}

where M$_{\rm AB}$ \cite{AB_magnitude} is defined in SI unit as\-

\begin{equation}
M_{AB}=-2.5\;\log_{10}(f_{\nu})-56.10\,\,\,\,\, f_{\nu}\,\mathrm{in}\,\mathrm{W}\,\mathrm{m}^{-2}\, \mathrm{Hz}^{-1} \label{eq:mABsi}
\end{equation}

The energy of the photons that the telescope
received can be expressed as a function of the flux
spectral density $f_{\nu}$. For the purpose of simulation, we approximated a flat spectrum in the corresponding wavelength band. By dividing this energy by the
the energy of a photon $h\nu_{eff}$, we can obtain the number of
photons collected by the telescope for one star as a
function of $f_{\nu}$, and then, as a function of $m_{AB}$

\begin{equation}
N_{\gamma}=\frac{E_{W}}{E_{\gamma}}=\frac{1}{h\,\nu_{eff}}\sideset{}{_{\nu_{1}}^{\nu_{2}}}\int f_{\nu}\,d\nu\,A\,T=\frac{1}{h\,\nu_{eff}}f_{\nu}\,\Delta\nu\,A\,T\label{eq:N_Photon}
\end{equation}

where A is the telescope’s collecting area, $\Delta\nu$ is the spectral bandwidth of observation, T is the exposure time. By substituting the $f_{\nu}$ expression from Equation. \ref{eq:mABsi} in Equation. \ref{eq:N_Photon}, we obtain

\begin{equation} \label{eq:number_of_photons}
N_{\gamma}=\frac{1}{h\,\nu_{eff}}\,A\,\Delta\nu\,T\,10^{\left(-\frac{m_{AB}+56.10}{2.5}\right)}
\end{equation}

Using Equation. \ref{eq:number_of_photons}, the number of photons corresponding to each star is calculated. A null flux map frame is generated and each star is added as a Poisson distribution. The thermal background photons are also added as Poisson distribution to this flux map.

\begin{figure}[h]
\begin{center}
\includegraphics[scale=.62]{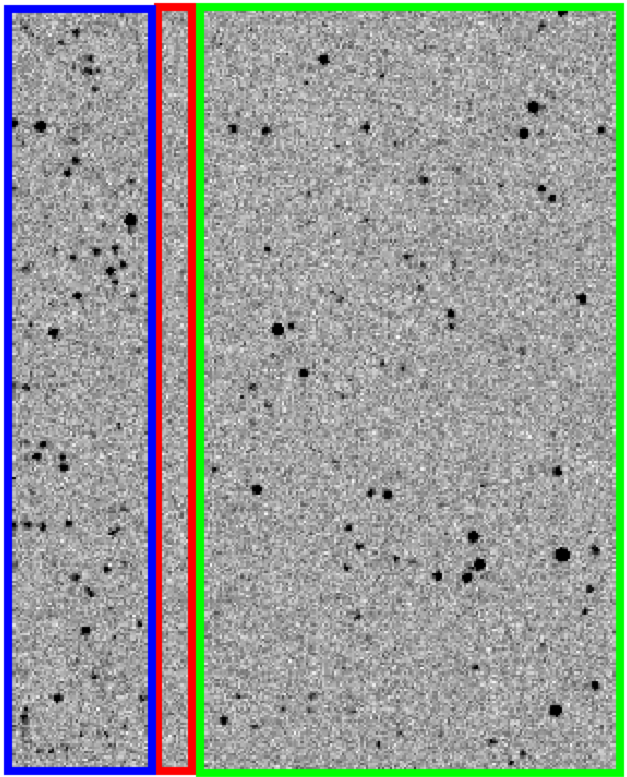}
\end{center}
\caption{An example of Emu single frame with exposure time $\sim$0.04 s using the 2MASS catalog at the Galactic anti-center. 
Here the blue, green and red box represents $J$, $W_{\rm E}$ bands and dead space between the bands respectively. }
\label{fig:Emu single fame}
\end{figure}

\begin{figure}[h]
\begin{center}
\includegraphics[scale=.55]{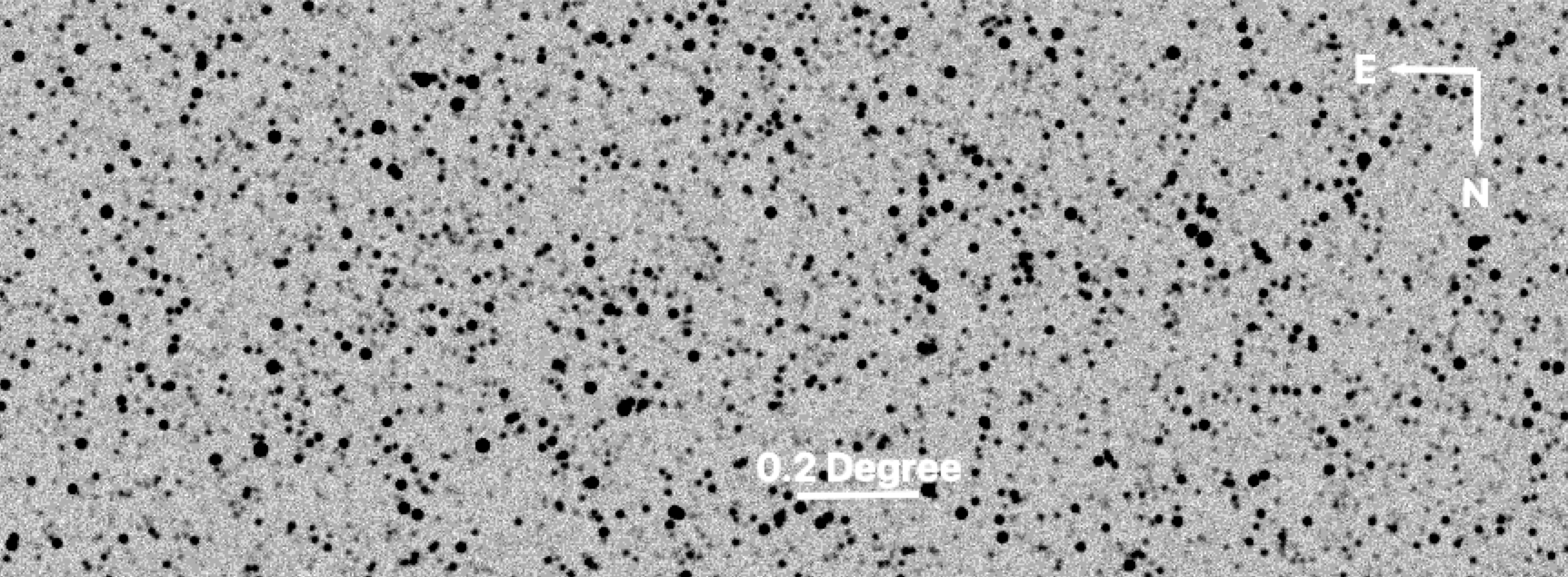}
\end{center}
\caption{The image shown is a Emu $W_{\rm E}$-band data simulation based on the current Emu baseline design and the 2MASS $H$-band input catalogue (extrapolated to the 1.4  $\mu$m  Emu W-band). This image was produced by shift and stacking $\sim$ 1200 frames, over one orbit pass at mid-latitude. The effective exposure time is individual pixel exposure time (0.04 seconds) $\times$ number of pixels in $W_{\rm E}$-band (180) = 7.2 seconds.} 


\label{fig:Emu W-band simulated image}
\end{figure}

\begin{figure}[h]
\begin{center}
\includegraphics[scale=.55]{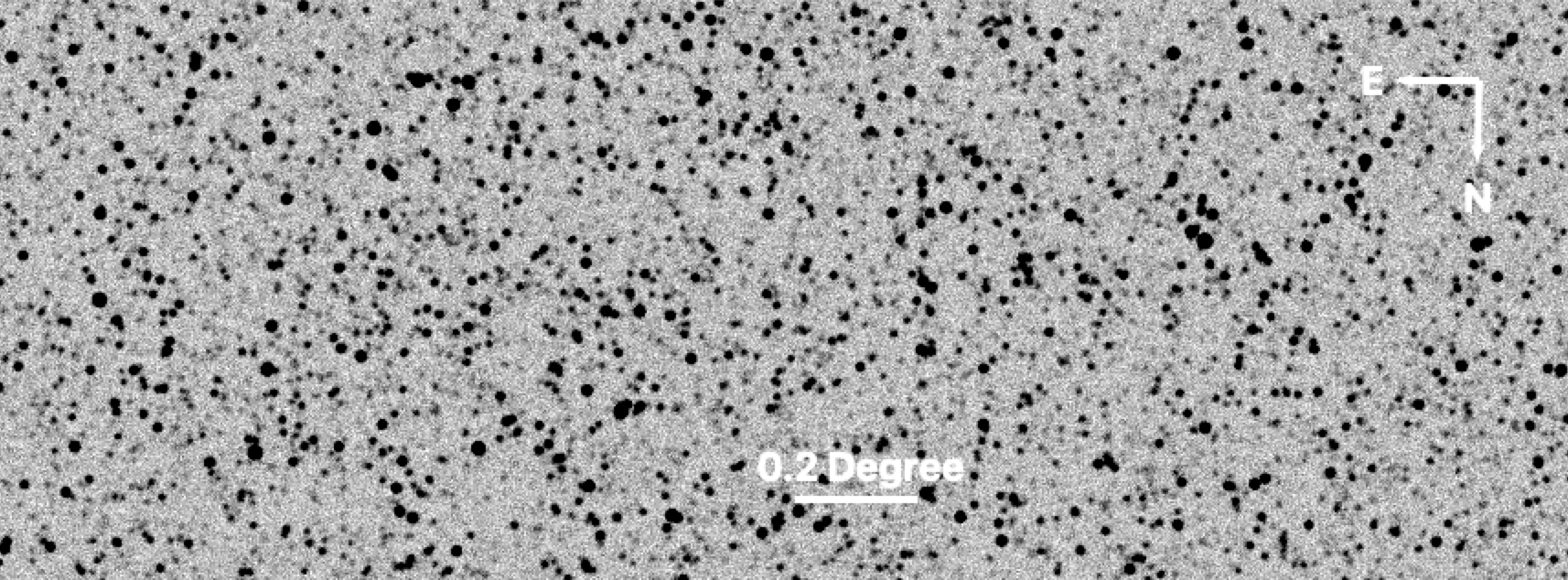}
\end{center}
\caption{The image shown is a Emu $J$-band simulation based on the current Emu baseline design and the 2MASS $H$-band input catalogue. This image was produced by shift and stacking $\sim$ 1200 frames, over one orbit pass at mid-latitude. The effective exposure time is individual pixel exposure time (0.04 seconds) $\times$ number of pixels in $J$-band (60) = 2.4 seconds.} 
\label{fig:Emu J-band simulated image}
\end{figure}

In order to find the relative shift between frames due to the vibrations on the ISS external platform, a cross-correlation between successive frames is performed. The offset obtained from this cross-correlation is used to correct for the relative shift between the frames. 
These $W_{\rm E}$-band and $J$-band frames are then shifted (each pixel by one column, based on the TDI technique), aligned, and co-added separately to produce a continuous strip with an effective exposure of the time taken for a target to cross the entire array. A weight map is also created in parallel corresponding to the number of shifts in order to correct the exposure gradient. The shift and the co-added frame are then divided with this white map to create the strip of the Emu star-field. An example of a single exposure simulated frame (with $J$ and $W_{\rm E}$ bands) is shown in Fig.\ref{fig:Emu single fame}. A simulated sky strip that Emu will deliver in $W_{\rm E}$ and $J$ bands are shown in Fig.\ref{fig:Emu W-band simulated image} and Fig.\ref{fig:Emu J-band simulated image} respectively.

\begin{figure}[h]
    
    \begin{center}
    \includegraphics[scale=.85]{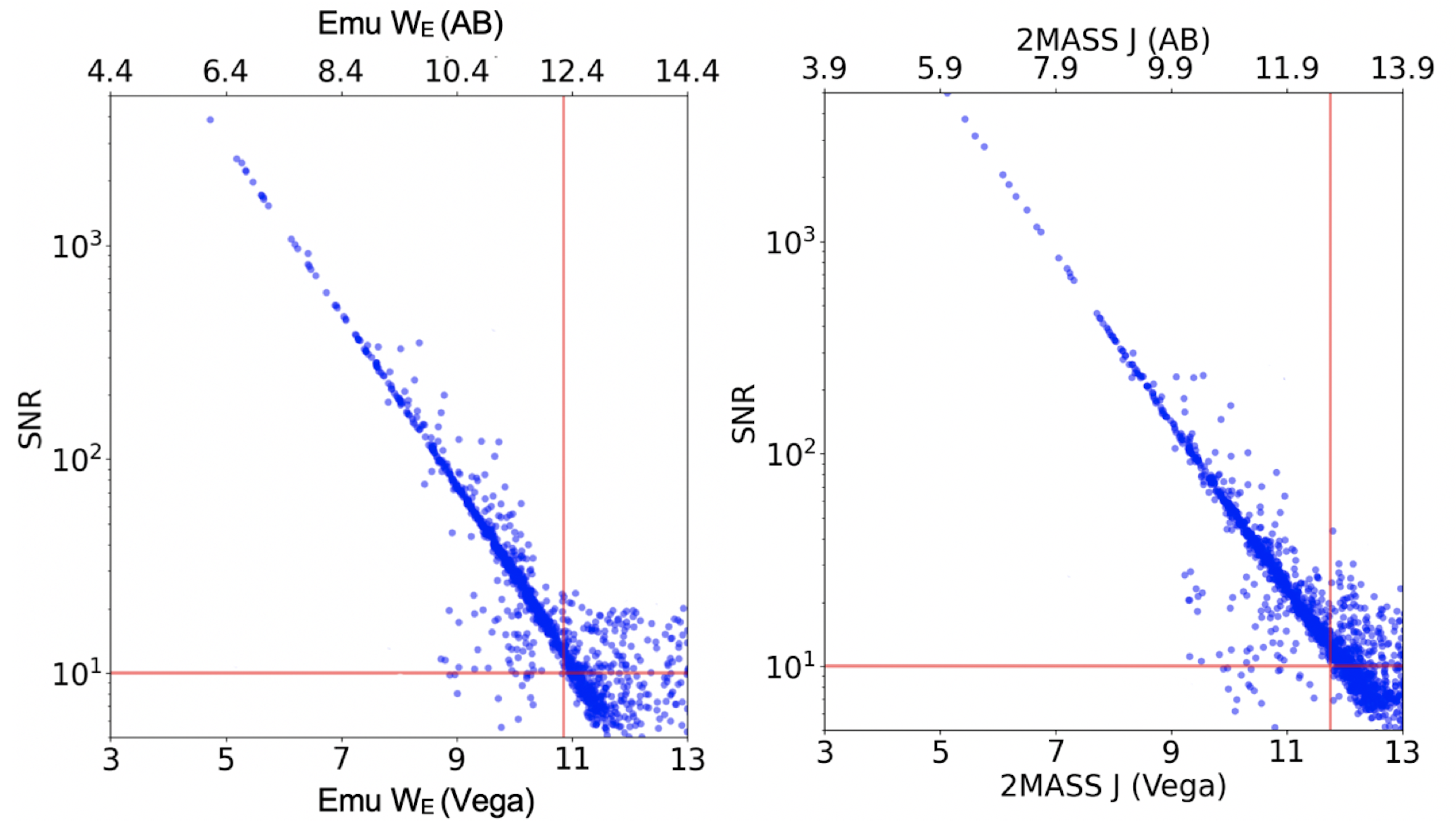}
    \end{center}
    \caption{{\it Left}: Emu $W_{\rm E}$-band sensitivity. To estimate Emu $W_{\rm E}$-band sensitivity we used 2MASS $H$-band data assuming a flat spectrum source. {\it Right}: Emu $J$-band sensitivity based on 2MASS $J$-band catalogue.}%
    \label{fig:Emu sensitivity}%
\end{figure}

To understand the sensitivity of the simulated data, we ran a source extractor algorithm\cite{Source_Extractor} on the simulated image. The source extractor algorithm extracts the flux value and signal to noise ratio (SNR) of each star in the simulated star-field. The extracted stars from the simulation are then cross-matched with the stars in the 2MASS catalog, to understand the sensitivity of the simulated star-field. Fig. \ref{fig:Emu sensitivity} shows the SNR of each star in the simulated sky strip. To estimate Emu $W_{\rm E}$-band sensitivity we used 2MASS $H$-band data assuming a flat spectrum source. It is clear from this figure that a limiting magnitude of $m_{AB}$ = $\sim$ 12.4 ($H$-band) and $m_{AB}$ = $\sim$ 12.6 ($J$-band) can be detected with an SNR of 10 with the Emu sky survey. The simulation shows that we can achieve the necessary sensitivity without a pointing and tracking system, by performing TDI from the ISS platform.

\subsection{Cosmic Ray Event Rate}
To understand the cosmic ray (CR) events experienced by Emu we have used the data from the Wide Field Camera 3 infrared channel (WFC3/IR) on the Hubble Space Telescope \cite{WFC3IR}. The WFC3 IR detector is a HgCdTe 1024 $\times$ 1024 array, with 18-micron pixels. 
As per the WFC3 instrument handbook \cite{2012wfci.book.....D}, the WFC3/IR camera has a CR event rate of 5 events/second/camera.
The WFC3/IR detector area is 3.39 cm$ ^{2}$ and this corresponds to 0.68 events/second/cm$^{2}$. For SAPHIRA, the detector area is 0.472 cm$ ^{2}$. And this will result in a CR event rate of 0.32 events/second for Emu. At 25 frames/second frame rate, this corresponds to 0.012 events/frames. This implies that the CR event rate has a little direct impact on Emu observations. However, the tolerance of SAPHIRA to such events is as yet unknown (indeed this is an important technical demonstration goal of the Emu mission), as are the implications for the control system electronics.

\section{Optimum pixel size for the mission}

For the Emu mission, a critical mission parameter is the selection of pixel size. A larger pixel size provides a wider field of view for a detector with a fixed number of pixels as well as leads to a greater TDI crossing time (hence increased integration time). This is important for sensitivity and frame to frame alignment. However larger pixel size would also increase the potential for source confusion (multiple sources in a single pixel). The optimum pixel size is depending upon the field alignment requirement for the single frame as well the confusion limit of the co-added frames.

\begin{figure}[h]%

\begin{center}
\includegraphics[scale=.65]{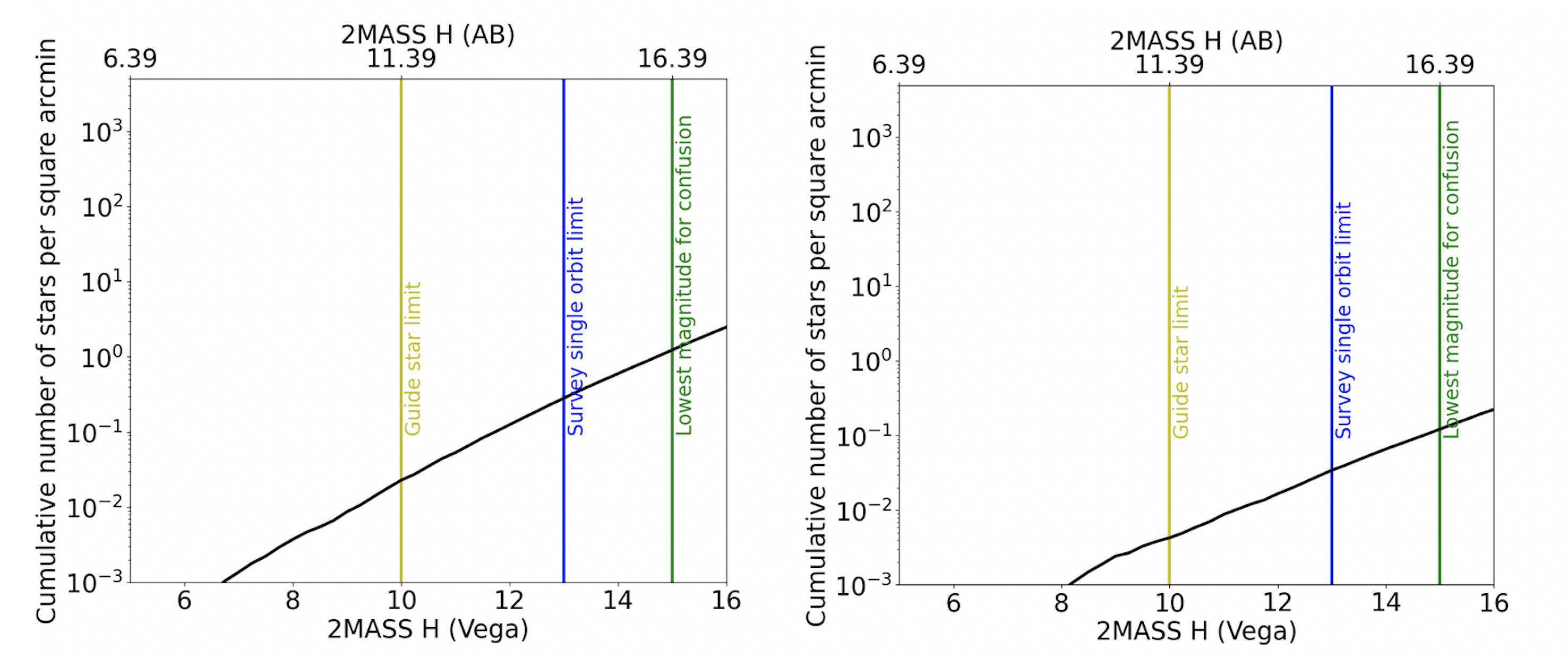}
\end{center}

    \caption{{\it Left}: Stellar density at the Galactic anti-centre.  {\it Right}: Stellar density at the Galactic North pole. The stellar density information is used to estimate the probability of more than three stars in Emu field of view and to estimate the confusion limit.}%
    \label{fig:Stellar density}%
\end{figure}

\begin{figure}%

\begin{center}
\includegraphics[scale=.75]{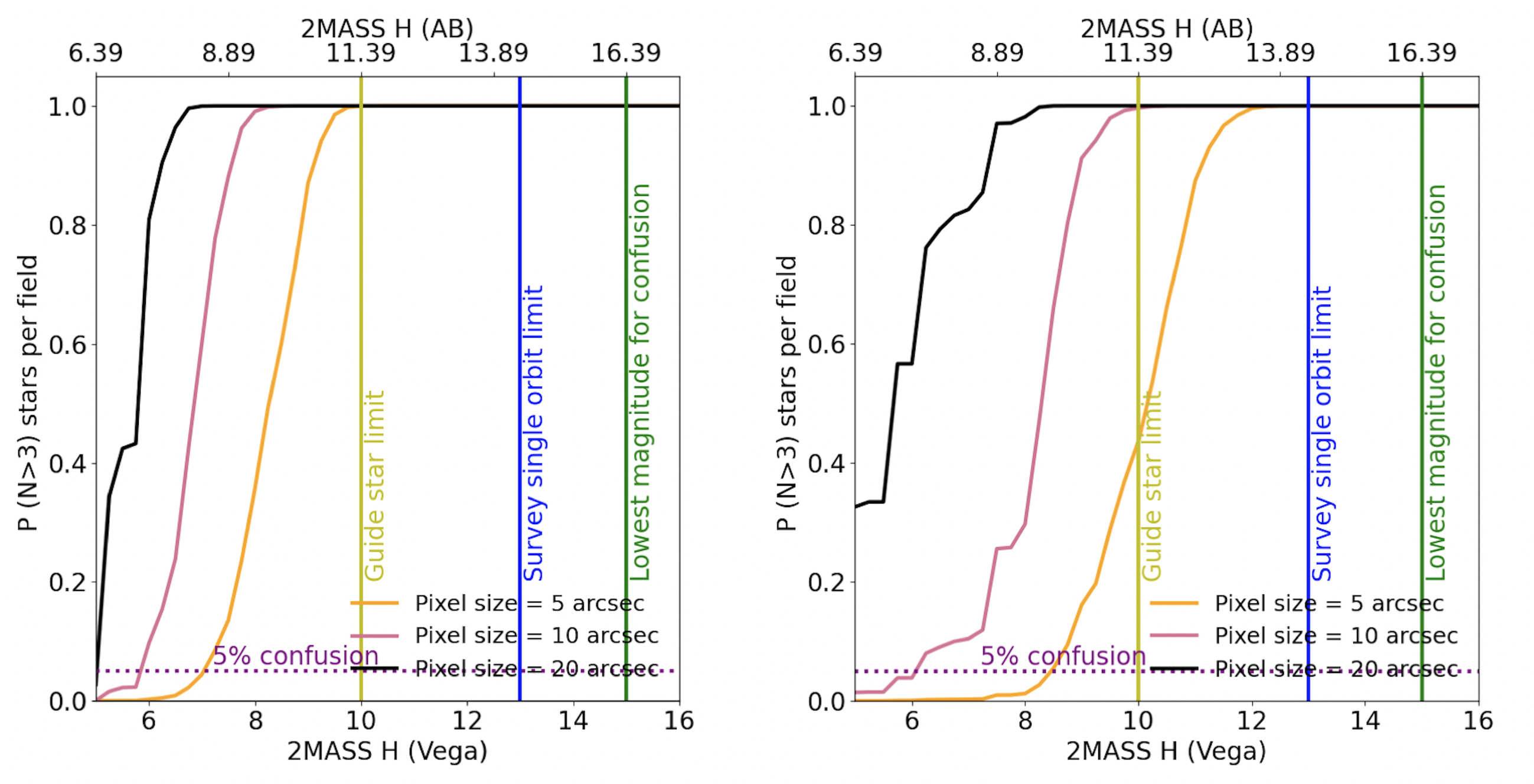}
\end{center}


    \caption{Probability of more than three stars in a single frame field of view at the Galactic anti-centre ({\it left}) and the Galactic North pole ({\it right}). A pixel size greater than or equal to $10^{\prime\prime} $ is required to satisfy the minimum 3 stars per frame condition for alignment purposes.}%
    \label{fig:PN3}%
\end{figure}

\begin{figure}%
\begin{center}
\includegraphics[scale=.65]{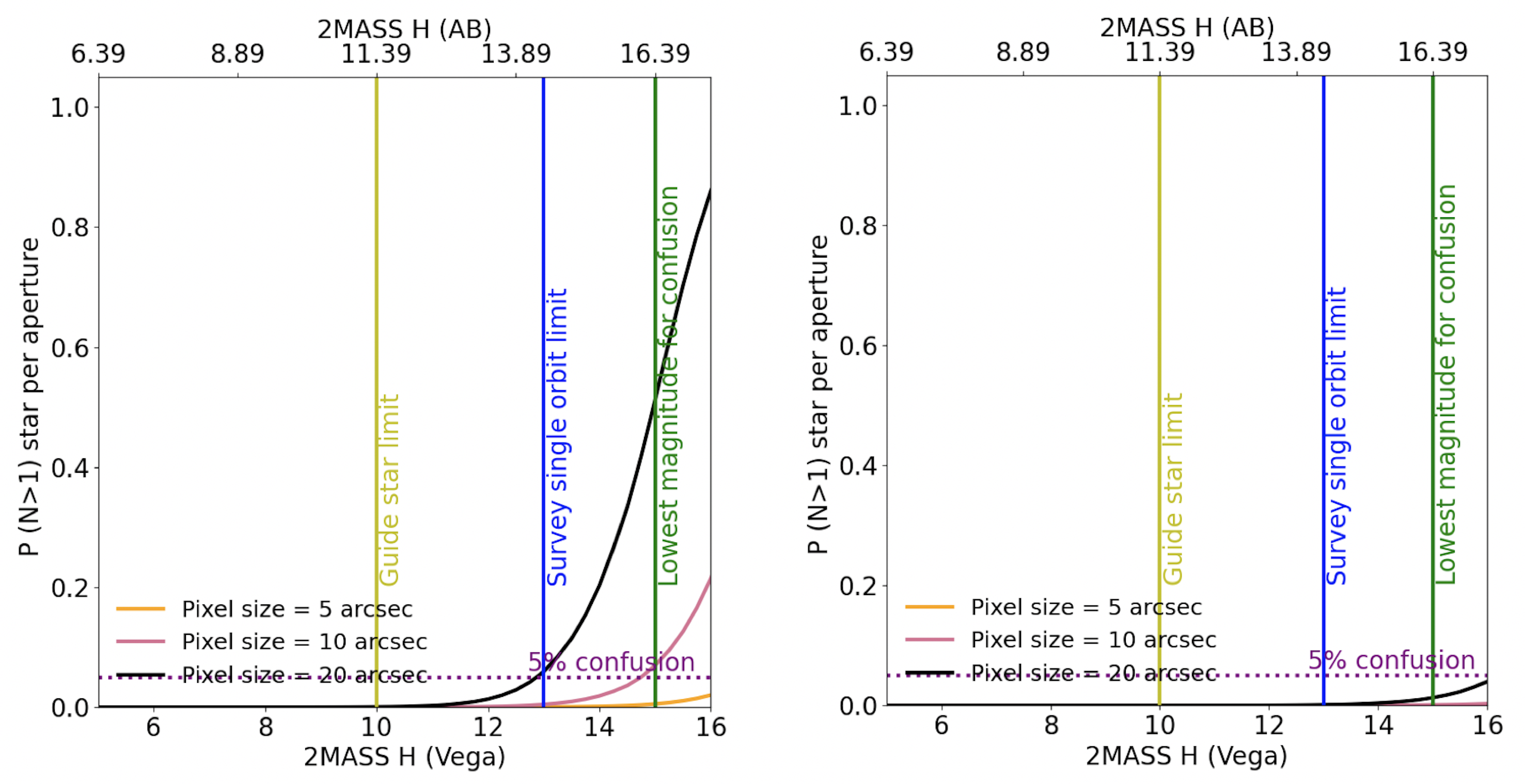}
\end{center}

    
    \caption{Probability of more than one star in photometric aperture at the Galactic anti-centre ({\it left}) and the Galactic North pole ({\it right}). With the $\sim$ $5^{\prime\prime}$ pixel, the probability of more than one star in photometric aperture is high at the Galactic anti-centre, a $10^{\prime\prime}$ pixel is optimal. }%
    \label{fig:PN1}%
\end{figure}

To perform alignment between different frames, it is required that every single frame has three or more guide stars detected to a significant level. The point spread function (PSF) of the final stack of the image ($PSF_{Out}$) should be blurred by no more than 10 \% of the input PSF ($PSF_{In}$) for an individual frame.


Centroid error in terms of Full Width Half Maximum (FWHM) and Signal to Noise Ratio (SNR) can be expressed as,

\begin{gather*}
\begin{aligned}
\sigma(centroid \; error)=\frac{FWHM}{SNR}\,,\\
\end{aligned}
\end{gather*}

and FWHM is,

\begin{gather*}
\begin{aligned}
FWHM= 2.35 \times \sigma(PSF_{In})
\end{aligned}
\end{gather*}

 So centroid error can be expressed as,

\begin{equation}
\begin{aligned}
\sigma(centroid \; error)= 2.35 \times \frac{\sigma(PSF_{In})}{SNR}\,,\\
\end{aligned}
\label{eq:centroid error 1}
\end{equation}
The output PSF would be 110 \% of input PSF and it can be expressed in terms of its standard deviation ($\sigma$) as,

\begin{gather*}
\begin{aligned}
{\sigma(PSF_{Out})}^{2}=1.1 \times{\sigma(PSF_{In})}^{2}\,,\\
\end{aligned}
\end{gather*}

And also,

\begin{gather*}
\begin{aligned}
{\sigma(PSF_{Out})}^{2}={\sigma(PSF_{In})}^{2} + {\sigma(centroid \; error)}^{2}\,,\\
\end{aligned}
\end{gather*}

i.e.

\begin{equation}
\begin{aligned}
{\sigma(centroid\; error)}^{2}=({1.1}^{2} -1) \times{\sigma(PSF_{In})}^{2}
\end{aligned}
\label{eq:centroid error 2}
\end{equation}

By substituting Equation. \ref{eq:centroid error 1} in Equation. \ref{eq:centroid error 2}, we will get,


\begin{gather*}
\begin{aligned}
SNR \sim 5
\end{aligned}
\end{gather*}

From this analysis, it can be estimated that a SNR of $\ge$ 5$\sigma$ is required to limit the output PSF blur to less than 10 \% of input PSF.

This implies guide stars' magnitude in every single frame should be at least 10 mag (2MASS $H$ (Vega)). So to Nyquist sample the image inputs, the full width half maximum (FWHM) is assumed as two pixels. In order to understand the best pixel size approximate for Emu, we have assumed a fiducial reference mission with three different pixels sizes, such as 5, 10, and 20 arcseconds. To estimate the optimum pixel size for the mission, the stellar density was derived empirically from representative fields at the Galactic anti-center and the Galactic pole as extremely high and low density cases. The Galactic center has a highly crowded stellar field, and the effect of source confusion is high compared to other regions of the sky\cite{Galactic_center_confusion}. To reduce the effect of source confusion we chose a sky field in the Galactic anti centre instead of the Galactic center. The stellar density estimated using 2MASS catalog data\cite{2MASS} at the Galactic-anti center and the Galactic north pole is shown in Fig.~\ref{fig:Stellar density}. The fiducial reference values such as the Emu pixel size and photometric aperture size (twice of FWHM) are also shown.

From the stellar density information, the Poisson probability of more than three stars in the field of view was estimated and it is shown in Fig.~\ref{fig:PN3}. It can be concluded that a pixel size greater than or equal to $10^{\prime\prime} $ is required to satisfy the – minimum 3 stars per frame- condition for alignment purposes. To estimate the confusion limit, the Poisson probability of more than one star in the photometric aperture was derived and this is shown in Fig.~\ref{fig:PN1}. From Fig.~\ref{fig:PN3} and Fig.~\ref{fig:PN1}, it can be concluded that a pixel size of $\sim$ $10^{\prime\prime} $ would be optimal for the Emu mission (when constrained to a single SAPHIRA focal plane array).

\section{The SAPHIRA electron Avalanche Photodiode Detector}

The Emu mission concept was developed based on the groundbreaking sensitivity of the SAPHIRA detector from Leonardo MW Ltd \cite{SAPHIRA_Baker}. The SAPHIRA adds an electron Avalanche Photodiode (eAPD) region below a traditional HgCdTe absorption layer, as part of a hybridized assembly with a CMOS readout integrated circuit (ROIC) with 32 parallel readout channels, each supporting a pixel rate of up to 10 MHz. The array has 320$\times$256 24 $\mu$m pixels. The detector is sensitive to a nominal wavelength range of 0.8–3.5 $\mu$m, with the effective quantum efficiency in the range 2.5–3.5$\mu$m depending on the selected avalanche gain\cite{Jamie_SAPHIRA}. A quantum efficiency of $>$ 60\% has been demonstrated for the SAPHIRA in Emu's wavelength band\cite{ESO_readnoise}.

\begin{figure}[h]
\begin{center}
\includegraphics[scale=0.8]{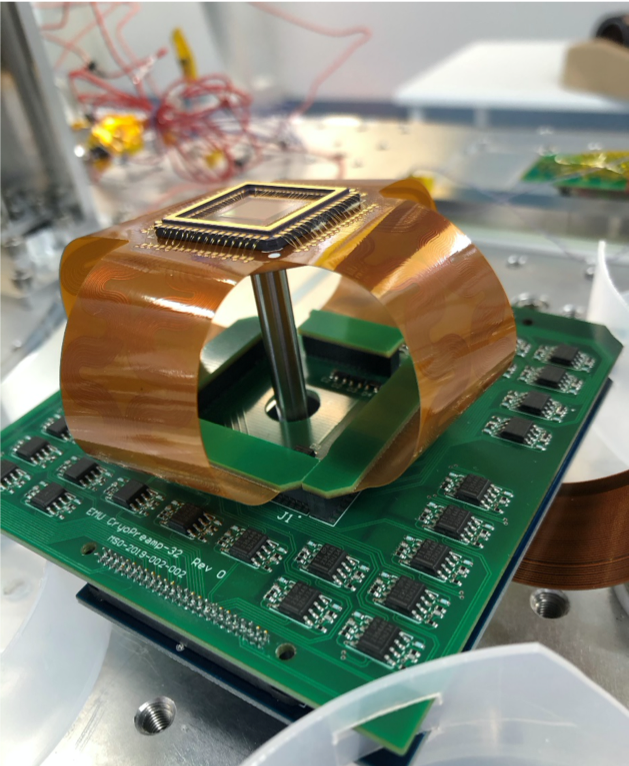}
\end{center}
\caption{The fully integrated SAPHIRA with flex cable and  cryogenic preamplifiers}
\label{fig:SAPHIRA}
\end{figure}
The avalanche gain is tunable via manipulation of bias voltages; and provides a near-noise-free signal gain of up to several hundred. This linear signal amplification results in an effective readout noise contribution of $\leq$ 1 electron per pixel per CDS image pair \cite{SAPHIRA_low_readout_noise}. This near-zero readout noise regime is critical to implementing the TDI-like observation mode for the Emu mission.

An avalanche gain of 45 was chosen for Emu to obtain a system readout noise $\le$ 1 $e^-$. Increasing the gain would reduce the dynamic range of the system. Emu’s bright magnitude limit is 4.5 m$_H$ (AB)  and this is basically set by the gain choice of 45. The gain can be controlled by setting the readout electronics reset voltage; thus, different modes of operation with different gain (thus different dynamic ranges) are possible.

SAPHIRA detectors are already used in ground-based astronomical instruments for fast wavefront sensing and fringe tracking applications\cite{SAPHIRA_Wavefront, Robo_AO_SAPHIRA, SAPHIRA_AO, SAPHIRA_interferometry}, but their utility in space applications is yet to be exploited. Specifically, their suitability for high-speed and/or photon-starved applications has profound implications for space-based near-infrared instruments, including TDI-like surveys of the sky from a Low Earth Orbit platform. This removes the need for a complex fully steerable (pointing and tracking) telescope that would add significant cost and risk to the project. This principle can also be applied to near-infrared (NIR) and short-wave infrared (SWIR) remote sensing Earth Observation applications.

The SAPHIRA detector has also undergone radiation testing, with a gamma and proton radiation campaign by the European Space Agency showing no permanent damage to the device\cite{SAPHIRA_radiation_testing}. A fully integrated SAPHIRA focal plane built at ANU is shown in Fig. \ref{fig:SAPHIRA}. One of the goals of the Emu mission is to advance the SAPHIRA detector and ANU `Rosella' control electronics to Technology Readiness Level 9 (TRL9). The use of the ISS as the orbital platform circumvents many of the engineering risks associated with small space missions as it removes the need for a full free-flying satellite bus, and offers features such as active cooling and payload return.

\section{Rosella detector control electronics}
Rosella is a high-performance detector controller for space, developed by ANU to address the lack of space-compatible pixel digitisation options for science grade optical sensors, which typically have no on-board analogue-to-digital converters (ADCs). 

\begin{figure}[h]
\begin{center}
\includegraphics[scale=1.2]{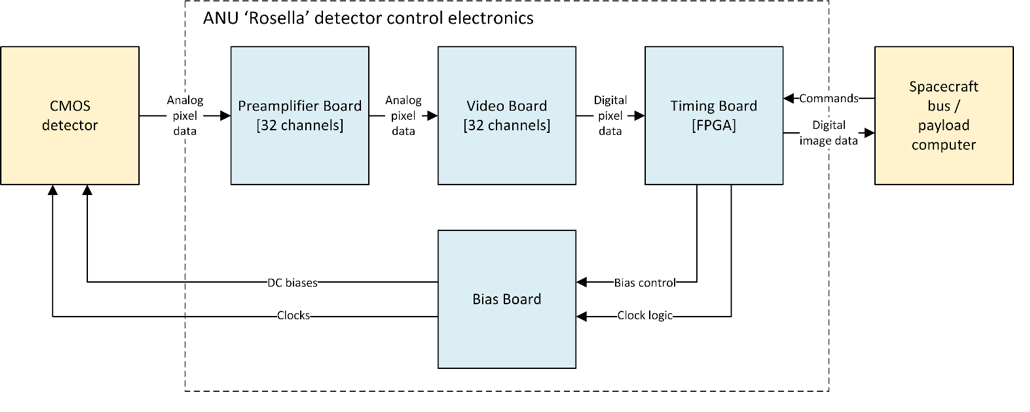}
\end{center}
\caption{Rosella electronics architecture}
\label{fig:Rosella}
\end{figure}

The Rosella system is compact and easily re-configurable for a wide range of visible and IR CMOS detectors including the Leonardo SAPHIRA and Teledyne HxRG family of infrared arrays. The Rosella architecture features a preamplifier board, bias board, video board, and a timing board based on a field programmable gate array (FPGA). The preamplifier board provides early amplification of the detector output signals very close to their origin\cite{preamplifier}, reducing the influence of electrical noise induced further down the signal chain. The bias board generates accurate and stable DC voltages for the detector, including the SAPHIRA's variable avalanche gain bias.

\begin{figure}[h]
\begin{center}
\includegraphics[scale=1.1]{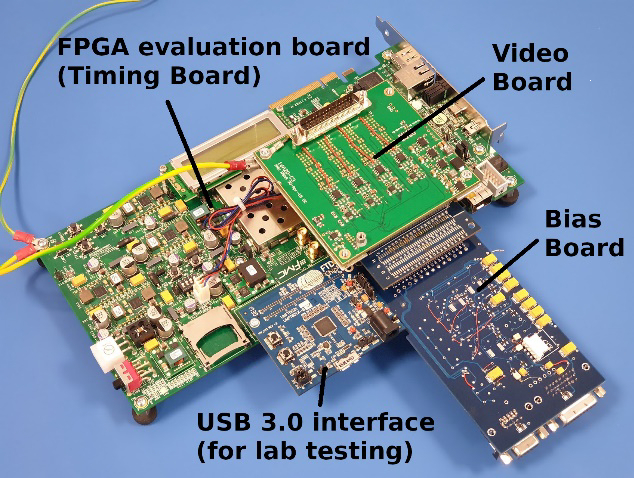}
\end{center}
\caption{Breadboard prototype of Rosella electronics including a commercial FPGA evaluation board, one of two 16-channel video boards for analog-to-digital pixel conversion, and a bias board for detector biases and clock conditioning}
\label{fig:Rosella prototype}
\end{figure}

\begin{figure}[h]
\begin{center}
\includegraphics[scale=0.75]{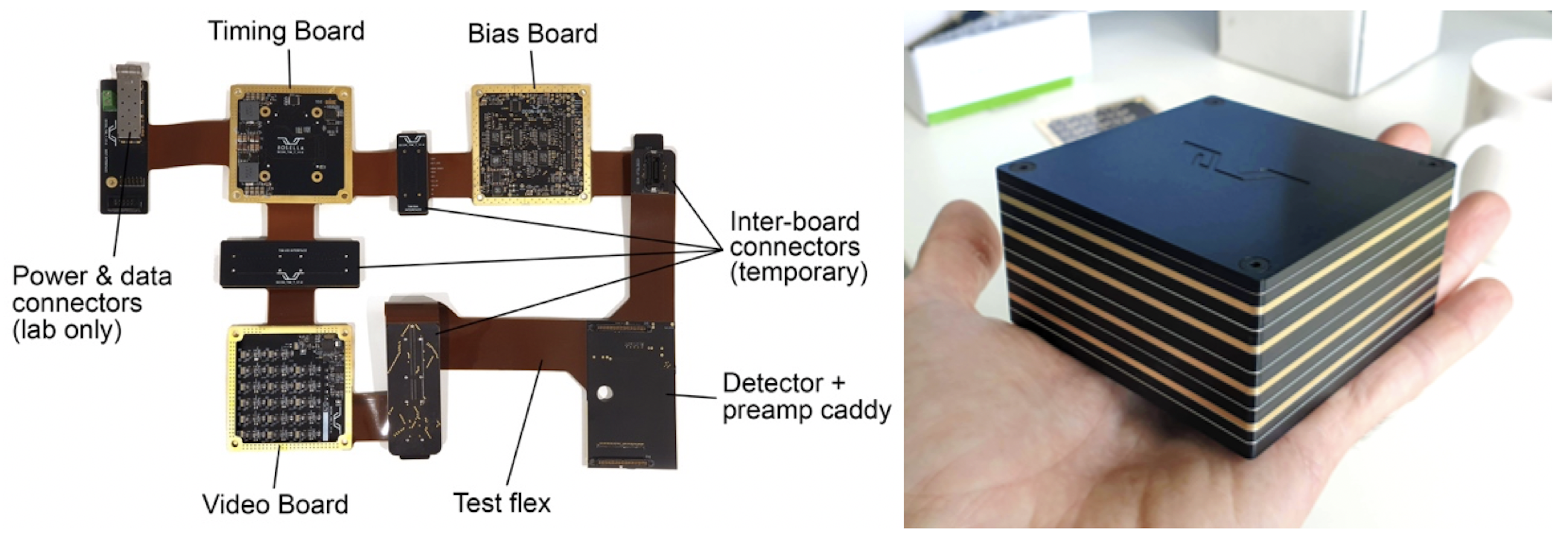}
\end{center}
\caption{{\it Left}: Rosella engineering model `FlatSat'; the inter-board connectors are temporary for testing, but will be a continuous flex connection on the flight model. {\it Right}: Rosella 0.5 U enclosure mock-up showing interleaved PCBs and aluminium enclosure walls; not shown are the inter-board flex circuit connections, which bend to allow folding and unfolding of the entire rigid-flex assembly.}
\label{fig:Rosella prototype}
\end{figure}

The video boards host 32 analog-to-digital converters (ADC) channels for digitising the incoming pixel stream. These devices are selected for low noise performance. The timing board manages the entire system, including bias configurations, clock pattern generation, ADC triggering, image processing, and communication with an external host computer through one of several standard protocols (e.g. Ethernet). Although the Emu mission requires a 50 Hz frame rate, the Rosella controller has been designed with high-resolution Earth Observation missions in mind, with frame rates approaching 1 kHz. A breadboard prototype of Rosella (Fig. \ref{fig:Rosella prototype}) has been successfully demonstrated on-sky at the ANU 2.3 m Telescope.

Rosella's final version will occupy a volume of approximately 0.5 U, comprising a connector-less printed circuit board (PCB) assembly based on rigid-flex technology. Each rigid PCB section features an outer thermal conduction region that interfaces with an aluminium wall that forms a contiguous and enclosed board stack by folding the flex circuit sections. The enclosure also provides light-tightness for payloads sensitive to infrared emission (thermal `glow'). Fig. \ref{fig:Rosella prototype} shows a `FlatSat' engineering model of Rosella and a mock-up of the PCB and enclosure assembly.

\section{Integrated Detector Cooler Assembly}
\label{IDCA}
\begin{figure}[h]
\begin{center}
\includegraphics[scale=0.7]{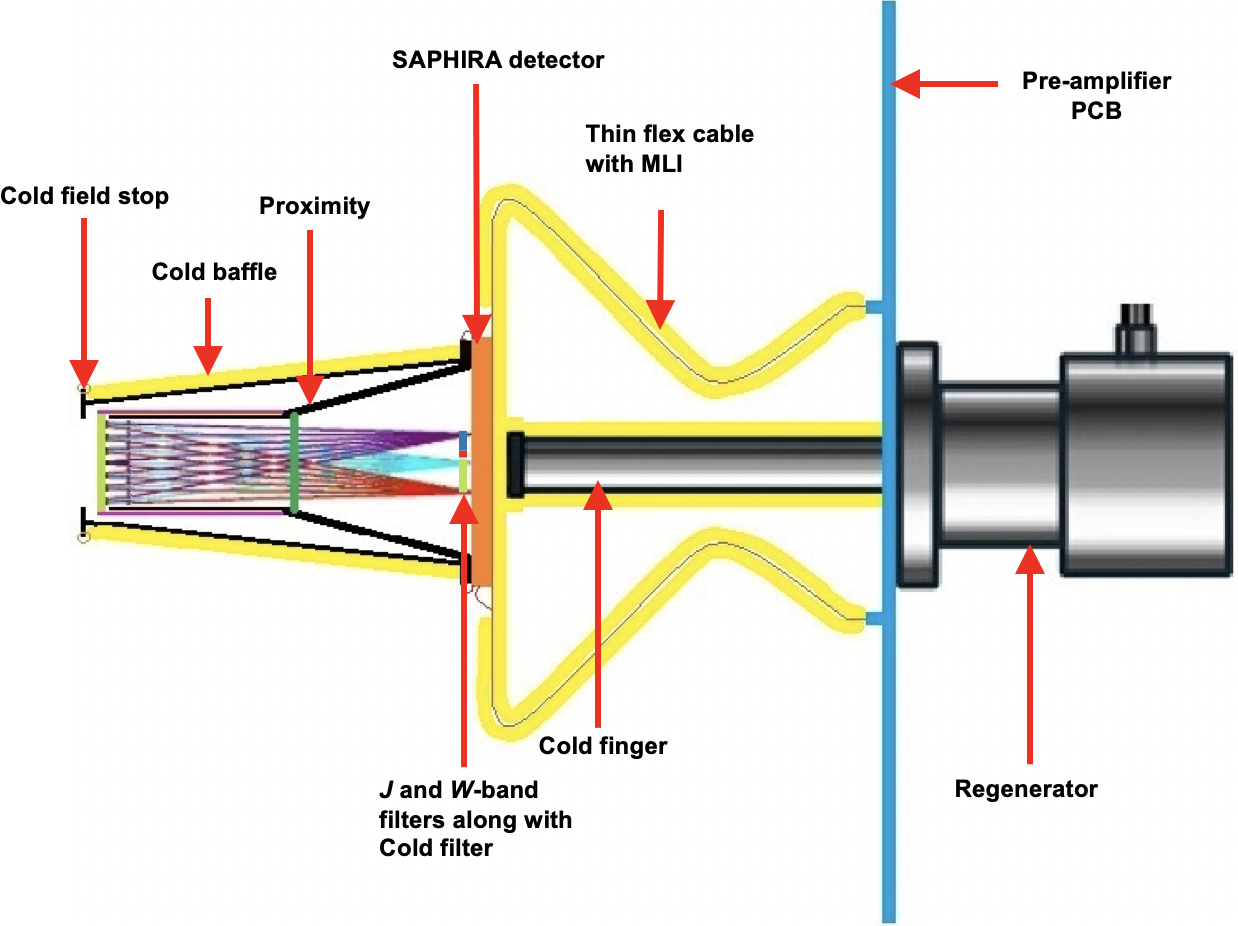}
\end{center}
\caption{IDCA and cold baffle concept layout.}
\label{fig:IDCA}
\end{figure}

The Integrated Detector Cooler Assembly (IDCA) system contains the cold optics, detector, cooling system, and mount. It is required to maintain the detector and cold optics at specified temperatures to meet the instrument performance requirements. 
The IDCA layout is shown in Fig. \ref{fig:IDCA}.
 
The assembly must also make provisions for trapping volatile contamination in order to avoid degradation of the performance of the cold optics and detector. The operating temperature requirement for the SAPHIRA detector is 80 K and this will be provided by a miniature split-Stirling cooler UP8497/01 from Thales Cryogenics. There is no Dewar, so the Stirling may work only when Emu is in a vacuum. This means Emu can be operated only under an external vacuum. Almost one hundred electrical wires (thin narrow tracks of the Flexible Cable) are coming from the 80 K zone to the detector control electronics operating at approximately room temperature. 
The Regenerator with the Cold Finger is a part of the Stirling cooler. The detector (80 K) and the Cold Baffle ($\sim$ 200 K) are mechanically held only by the Cold Finger and don’t touch any other constructions. The Rigid-Flex PCB includes the Thin Flexible Cable (with 4 wings) and connects electrically SAPHIRA with the Pre-Amplifier PCB. The Thin Flexible Cable has low thermo-conductivity so the total heat load to the Cold Finger’s tip is within our miniature Stirling’s possibility to lift heat. The metal cold baffle is a truncated 4-side pyramid glued to the rim of the SAPHIRA package. At the tip of the Cold Finger, there are detector, Cold Baffle, Cold Filter and mounts. The rest of the Emu subsystems will operate at room temperature. The total mass of this assembly is around 10 grams. We have performed vibration tests with a dummy Cold Finger and dummy mass to be sure that the Cold Finger of our Stirling is strong enough not to crack during launch vibrations. These vibration tests have been performed at the National Space Test Facility, at the Mount Stromlo Campus of the Australian National University (ANU). The qualification test was based on General Environmental Verification Specification (GEVS) standard and yielded compliant results. The inner truncated 4-side pyramid (metal) holds the Cold Filter and the Teflon tube. This pyramid is glued to the SAPHIRA’s rim and gets cold from it. The outer pyramid’s base (metal) is glued to the base of the inner one.

\section{Optics}

\begin{figure}[h]
\begin{center}
\includegraphics[scale=1.3]{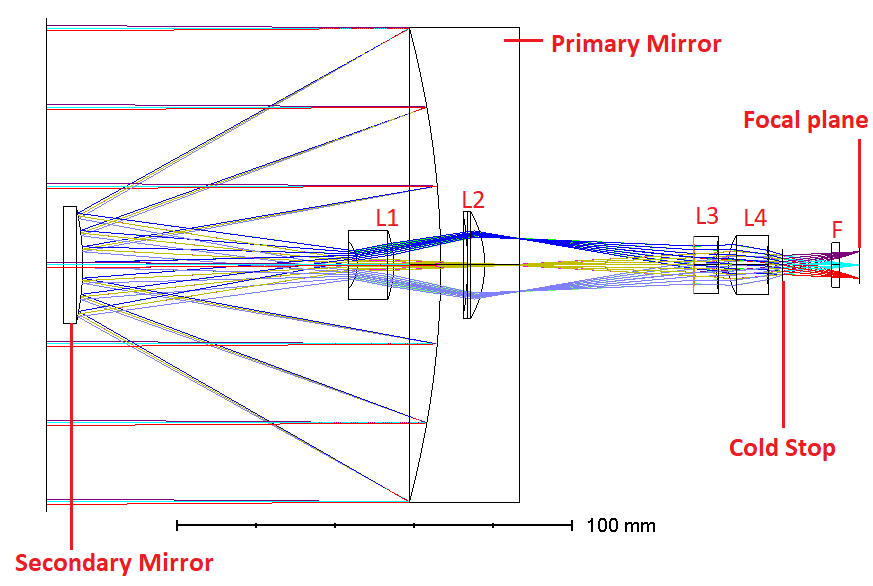}
\end{center}
\caption{The optical layout of the Emu payload. The primary mirror and secondary mirror together act as a Cassegrain telescope and its focus is reimaged by the relay lenses L1, L2, L3, and L4 to form a reimaged pupil at the cold stop. The science filters along with a cold filter (F) close to the detector focal plane will provide the required passbands for Emu.}
\label{fig:Optical Layout}
\end{figure}

Emu optics have been designed such that they will have control over the thermal background and will block the unwanted radiations from all directions, except for light passing through the telescope pupil. This is achieved by re-imaging the telescope's input pupil at an intermediate pupil plane cold stop. A cold aperture placed in this pupil image will block all the unwanted radiation not coming through the telescope. A set of four relay lenses (Anti-Reflection coated) are used to reimage the pupil to the cold stop and subsequently focus the light rays on the detector focal plane. A cold filter along with a science filter close to the detector focal plane will provide the necessary transmission and suppression for Emu. A Deuterated Potassium Dihydrogen Phosphate (DKDP) based filter is used as a cold filter to block long-wavelength thermal emission (see Section. \ref{Thermal emission analysis} for details). A ray trace of the optical system is shown in Fig.~\ref{fig:Optical Layout}.

\begin{figure}[h]
\begin{center}
\includegraphics[scale=0.9]{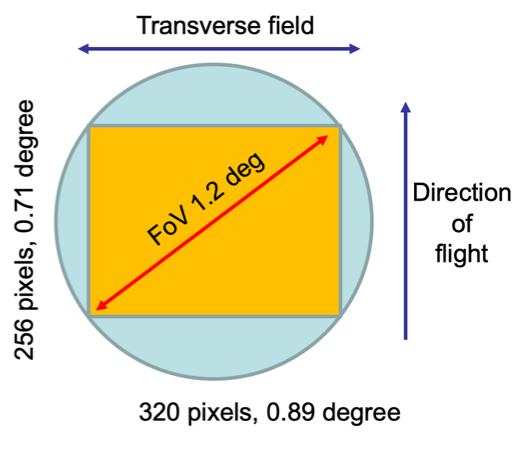}
\end{center}
\caption{The Emu FOV. The direction of flight is along the short FOV axis, with a diagonal FOV of $\sim$ 1.2 degree}
\label{fig:Emu FOV}
\end{figure}

The entrance aperture of the optical system is an 83 $\times$ 83 mm and the system focal length is 495 mm, yielding a plate scale of $ 10^{\prime\prime} $/pixel (at the 24 $\mu$m pixel pitch of the SAPHIRA eAPD detector). The diagonal FoV is $\sim$ $1.2^{\circ}$ (Fig.~\ref{fig:Emu FOV}). The mirrors will be made from low thermal expansion Zerodur material and will be coated with Gold for maximum reflectivity at 1.4 $\mu$m. The field lenses are of Zinc Selenide (ZnSe) and Zinc Sulfide (ZnS) materials. The spatial resolution requirement of the system is 2 pixels FWHM. The distortion of the optical system is less than 0.1 \%. The axial length of the optical system is around 190 mm, and this restriction is imposed by the 2U volume requirement on the optical unit. The throughput requirement of the optical system is $ \sim$ 80 \%. 

\section{Thermal emission analysis} \label{Thermal emission analysis}
 Emu will be operating in $W_{\rm E}$ and $J$ bands and it will be sensitive to the thermal emission from other parts of the instrument. This can be further complicated by two factors: the use of the SAPHIRA detector which has significant long-wavelength quantum efficiency in some modes of operation; and, the likely high telescope operating temperature ($\leq$ 50 $^{\circ}$C). This results in a careful balance of operating requirements for the Emu system. We have developed a thermal model to understand the various contribution of thermal noise to the detector plane. 
 
The SAPHIRA eAPD detector was originally developed for high-speed wavefront sensing with adaptive optics systems, An application in which dark current is not a major concern due to the high frame rate.  While device characteristic has improved in later implementation of the Leonardo eAPD detector family, the SAPHIRA detector available to the Emu project in the early phase has a relatively high dark current of $\sim21$\, $e^-$\,s$^{-1}$, or approximately 1 electron per Emu frame (for the flight version we will use SAPHIRA with lower dark current $\sim8.4$\, $e^-$\,s$^{-1}$). So for the thermal analysis, a worst-case scenario of dark $\sim21$\, $e^-$\,s$^{-1}$ current is assumed. This sets a fiducial reference against which to develop the allowable thermal background noise budget for Emu. It is comparable to the assumed effective readout noise of $\approx1$\,electron per pixel derived from selecting an avalanche gain setting that maintains sufficient dynamic range while still suppressing intrinsic readout noise.

The thermal model for Emu assumes an uncooled telescope structure at an equilibrium reference temperate of 50$^\circ$\,C. The combined warm optical train is assumed to have an effective emissivity of $\epsilon \le$ 0.15 (based on commercial grade infrared gold coatings for the two telescope mirrors and anti-reflection coatings on each surface of the four lens elements. Science and short-pass blocking filters are cooled to 150\,K resulting in negligible thermal emission in the $J$ and $W_{\rm E}$ bands. There is an additional contribution (assumed unit emissivity) from the central obstruction which occupies 12.5 \% of the pupil in the current configuration. The current optical design includes baffling for the central obstruction in the cold stop structure, ensuring a lower emission temperature for this component. This component is also cooled to 150\,K and hence is also negligible when fully masked. The thermal background provides a hard lower limit on the thermal background (with equilibrium temperature and effective emissivity dictated by external design considerations for Emu).

\begin{figure}[h]
\begin{center}
\includegraphics[scale=0.5]{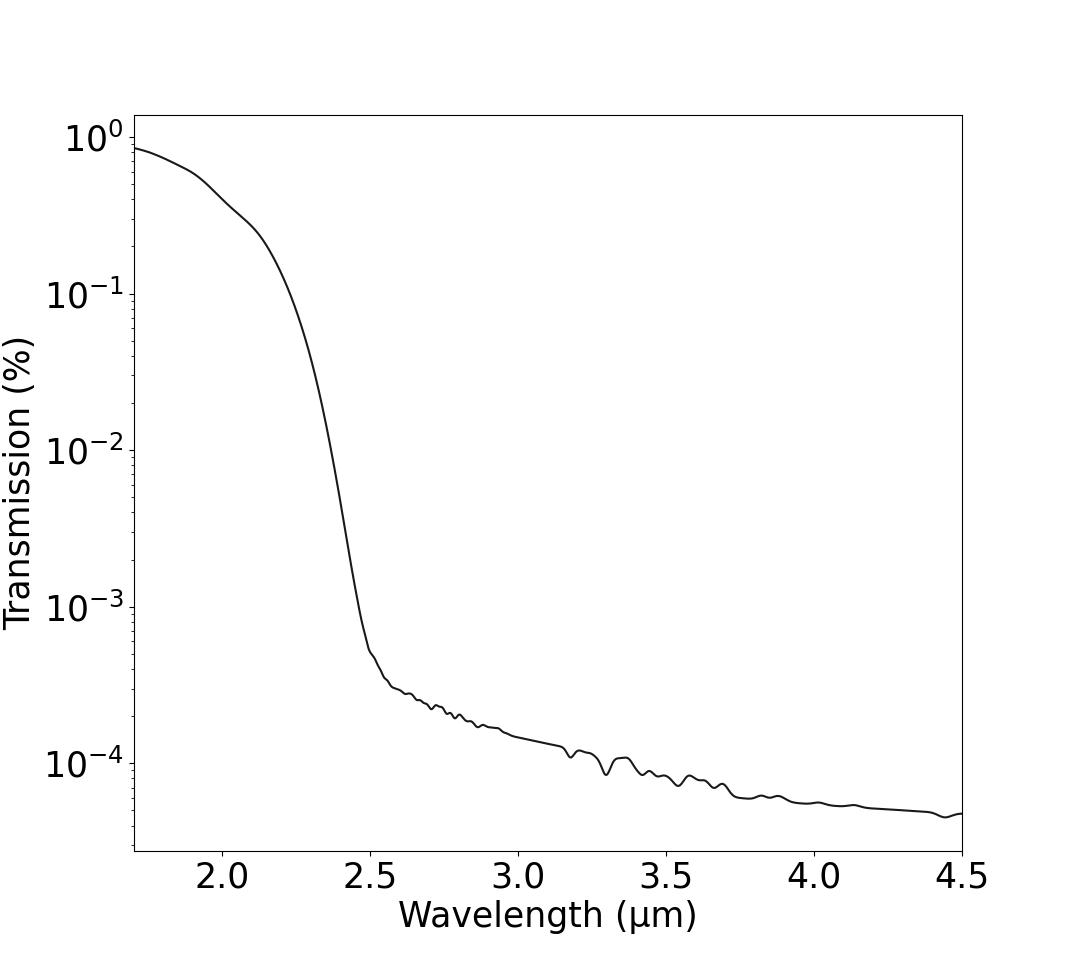}
\end{center}
\caption{The measured DKDP filter transmission profile. It has high short-wavelength transmission, and has good suppression beyond 2.5 \,$\mu$m, which is required to realise Emu pass bands.}
\label{fig:DKDP Transmission}
\end{figure}

The Emu optical design includes an intermediate pupil image and associated optical stop. As discussed earlier  in Section \ref{IDCA} and Fig. \ref{fig:IDCA}, this cold stop is part of the cooled detector cold baffle, and hence blocks thermal radiation from outside the telescope pupil reaching the detector. The Cold stop modeled to achieve an equilibrium temperature of 200 K. It remains undecided as to whether the cold stop will be provided with occluding structure to mask the Central obstruction, as the performance gain would be marginal due to the current analysis.

A cold short-pass filter is required to block long-wavelength thermal emission (from the detector enclosure and from any low transmission red leak associated with the science filters). The design of conventional interference filters for this role can be challenging due to the possibility of the same transmission overtones the filter is designed to suppress in the primary science filer. A common practice in infrared instrumentation has been to use the crystalline material DKDP. DKDP has high short-wavelength transmission, and  fine-tuning the level of deuteration, functionally zero transmission beyond 2.5\,$\mu$m can be achieved with compromised performance in the transition region 1.8-2.5\,$\mu$m, see Fig. \ref{fig:DKDP Transmission} (DKDP filter transmission measurement shown here was done at the ANU Research School of Physics using the FTIR spectrophotmeter in vacuum mode). The DKDP filter will be mounted close to the detector (bonded to the detector carrier plate) to provide wide angle suppression of thermal emission. This optical arrangement is predicted to result in a thermal background of $\le$ 7.5 $e^-$ s$^{-1}$ per pixel, of $\approx 0.3$ $e^-$ per Emu frame at 25\,Hz, with the dominant source of thermal background being the direct thermal emission from the optical system in the longer wavelength Emu $W_{\rm E}$-band. This will require experimental verification as is it remains a critical performance risk for the Emu system sensitivity.
 
\section{Observation Strategy} \label{Observation Strategy}

\begin{figure}[h]
\begin{center}
\includegraphics[scale=0.42]{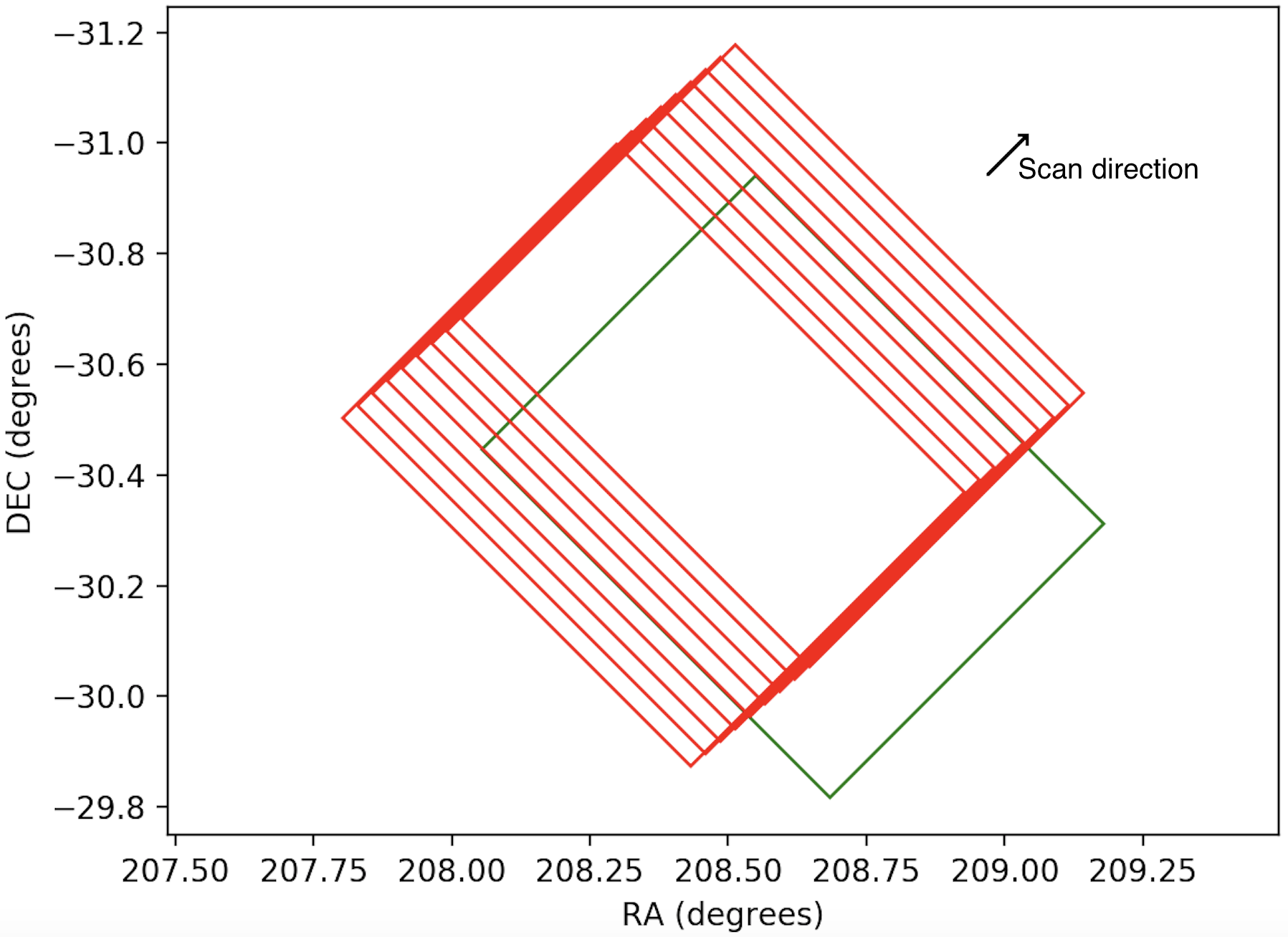}
\end{center}
\caption{The field overlap between successive orbits is shown. The green box corresponds to the field of view in initial orbit and red boxes corresponds to the field of view in the successive orbits with 0.5 s time difference}
\label{fig:field_overlap}
\end{figure}

The ISS provides a novel platform for the Emu payload to perform sky surveys in the water absorption band. A relatively wide field zenith-looking telescope with TDI capability can perform the Near IR sky survey without active pointing. The detector clock has to be synchronized with the transit speed of ISS. There is a significant field overlap ($\sim$ 74 \%) between observations in adjacent orbits. This would enable to shift and stack multiple orbits of data, thereby increasing the sensitivity of the final data product. The field overlap is shown in Fig.\ref{fig:field_overlap}, the arrow indicates the moving direction of ISS.

Emu’s commissioning begins soon after Emu is mounted on the ISS platform. The duration of the commissioning period is 4 weeks and it will generate raw data of ~7.5 TB. In the first week of commissioning, Emu outgasses itself and the detector won't be operated. All data during this period will be downloaded. In the next 3 weeks, the normal operation will start. 
The main science operation will begin after the commissioning period. The total data generated during this period will be around 55 TB and with at least one additional data downlink session/week, 0.3\% of the data can be downlinked. Emu will perform the astronomical observations only in the dark portion of the ISS orbit. In the default mode, it will acquire images at 50 Hz NDR, (25 Hz CDS pairs) and will store the data in the onboard solid-state memory. The raw data rate of Emu is 66 Mbits/sec and this corresponds to $>$300 GB per day. The available downlink rate is around 1 Mbits/sec and with only one uplink/downlink session per week available, for 8 hours. Emu has an onboard storage module with 60 TB of solid-state memory, to store the data corresponding to six months of operation. With the available downlink capability, we will be able to downlink only 1\% of this data to Earth for the purpose of quality control and monitoring. The solid-state memory will be returned to Earth for post-processing when Emu’s 6-month mission is completed.

\section{TDI-like imaging in the context of GaiaNIR mission} \label{GaiaNIR mission concept}

GaiaNIR is a new all-sky near IR astrometry mission under consideration by the European Space Agency (ESA) as a part of its Voyage 2050 future science mission program. One of the significant technical challenges for GaiaNIR is the requirement for low noise NIR detectors which can perform TDI-like imaging\cite{Gaia_white_paper}.

\begin{figure}[h]
\begin{center}
\includegraphics[scale=0.635]{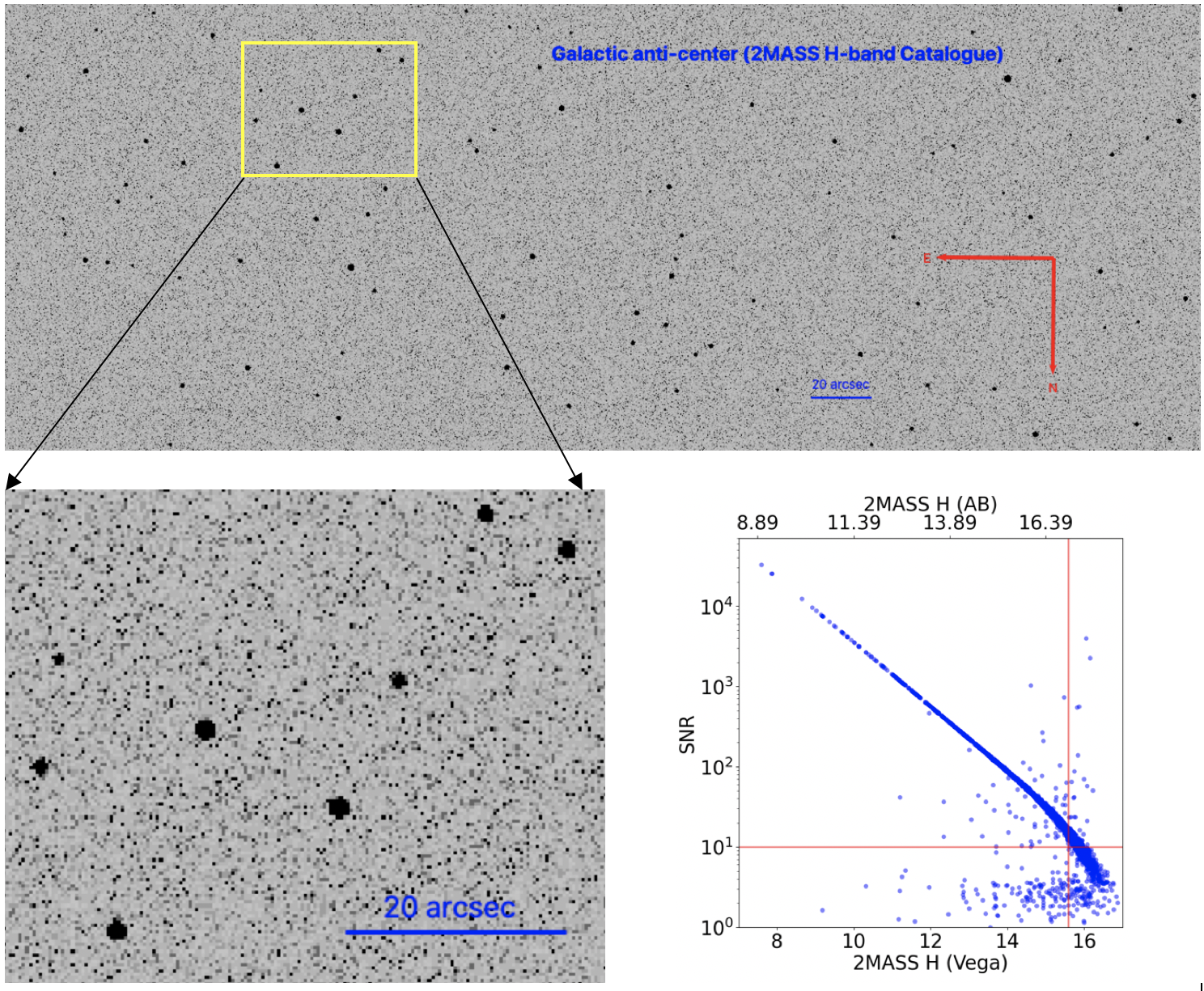}
\end{center}
\caption{
The image shown is a simulation for GaiaNIR like mission based on TDI concept.  Star field at the Galactic-anti center was simulated based on the the 2MASS H-band input catalogue. Also shown is the simulated GaiaNIR concept simulation signal-to-noise model.}
\label{fig:GAIAI_NIR}
\end{figure}

Emu will be performing a TDI-like NIR sky survey using the Leonardo SAPHIRA eAPD 320 X 256 detector array. Leonardo MW Ltd. is developing large format 2K x 2K eAPD detector arrays and could be a potential detector for the GaiaNIR mission. The metallicity information derived from the proposed Emu $W$-band filter will be useful in interpreting the GaiaNIR data; since no onboard spectroscopy is planned in the latter mission. The Emu simulation suite as outlined in section \ref{Simulation}, has been modified, to present TDI-like imaging simulations relevant to the GaiaNIR mission concept. A single 2K x 2K detector, with a pixel sampling scale 0.3$^{\prime\prime}$ per pixel and an exposure time of 3.75 milliseconds per frame is assumed for this simulation. The simulation in Fig. \ref{fig:GAIAI_NIR} is  not an actual representative of a the GaiaIR mission, but it shows that the Emu simulation can be modified by changing the input parameters to analyse the sensitivity of TDI-like GAIA NIR mission based on low noise SAPHIRA like eAPD detector arrays.

\section{Summary}
TDI-like imaging has great potential for near-infrared observations from space, and this is possible with the low-readout noise and high-speed SAPHIRA detectors. Emu is a proposed mission on the ISS and will demonstrate this key capability in space along with performing an infrared sky survey in  a novel $W_{\rm E}$-band at 1.4 $\mu$m, inaccessible from the ground.
The Emu survey will provide a unique new photometric
measurement which, when properly combined in a statistical sense with the full suite of contemporary data, will allow a wide range of questions in stellar atmospheric
research to be addressed. The measurement of the critical oxygen-to-hydrogen abundance of both dwarf and giant cool-stars (as probed by their water absorption band)
will provide critical new empirical constraints to state-of-the-art astrophysical models. It has the potential to advance not only our observational understanding of the chemodynamical evolution of our Galaxy through the study of its coolest--and most common--stars and their planets, but will also facilitate the development of state-of-the-art stellar atmosphere models.
Combined with the
revolutionary data released from missions such as the European Space Agency's (ESA) Gaia survey, this work enables the powerful
development of stellar atmosphere modeling. Also, this mission will be the first in-orbit demonstration of SAPHIRA, elevating its space-readiness to TRL9 along with our Rosella readout electronics. The Emu W-band survey mission will be the first science program to
demonstrate the power of these new devices for astronomy and science in general. The Emu mission passed its Conceptual Design Review in May 2019 and prototyping of critical and high-risk systems is underway.

\acknowledgments 
The authors pay their respect to the traditional and rightful custodians of the land on which this project takes place: the Ngunnawal and Ngambri peoples. We extend our respect to all First Nation people, their continuing resilience, culture and contribution. A significant funding for the development of the Rosella electronics system for Emu was provided by ANU InSpace. Thanks to Patrick Kluth and Lan Fu at the ANU Research School of Physics for assisting in the DKDP filter transmission measurement. This work makes use of data products from the Two Micron All Sky Survey (2MASS), which is a joint project of the University of Massachusetts and the Infrared Processing and Analysis Center/California Institute of Technology, funded by the National Aeronautics and Space Administration and the National Science Foundation. This work made use of Astropy\cite{Astropy}, SciPy\cite{Scipy} and NumPy\cite{NumPy} python packages.
LC is the recipient of an ARC Future Fellowship (project number FT160100402). M{\v Z} acknowledges funding from the Australian Research Council (grant DP170102233) and support from the Consejer{\'{\i}}a de Econom{\'{\i}}a, Conocimiento y Empleo del Gobierno de Canarias and the European Regional Development Fund (ERDF) under grant with reference PROID2020010052.

\pagebreak
\section{Appendix}\label{Appendix}

\bibliography{report} 
\bibliographystyle{spiejour}

\vspace{1ex}

\noindent \textbf{Dr Joice Mathew} is an instrumentation scientist at the Australian National University (ANU). His research interests include electro-optical payload development, UV and IR instrumentation, systems engineering, space instrumentation, and qualification. Joice obtained his Ph.D. in astronomical space instrumentation from the Indian Institute of Astrophysics, Bangalore. Before joining ANU, he worked as an instrument scientist for the Solar Orbiter mission at the Max Planck Institute for Solar System Research, Germany. 

\noindent \textbf{Dr James Gilbert} leads the optical detector program at the ANU Research School of Astronomy and Astrophysics, where his group specialises in ground and space-based focal planes and control systems. He holds a DPhil in astrophysics from the University of Oxford and an MEng in Electronics with Space Science and Technology from the University of Bath.

\noindent \textbf{Prof Robert Sharp} is an astronomer with over 20 years of experience in delivering instrumentation for infrared, integral field spectroscopy and multi-object spectroscopy. He is the principal investigator for Giant Magellan Telescope Integral Field Spectrograph, GMTIFS.

\noindent \textbf{Alexey Grigoriev} got his Master of Science in Astronomy degree from the Department of Physics, Lomonosov Moscow State University, in 1978.  He worked at Russian Space Research Institute for more than 40 years, creating scientific space spectrometers (UV to thermal IR). 
They operated on Venus's surface, near Comet Halley, and in Martian orbits. He is the technical chief of Fourier-spectrometer TIRVIM (Exo-Mars mission).
Since 2018 he works as Project Engineer at Australian National University.

\noindent \textbf{Adam Rains} completed his PhD in astronomy and astrophysics at the Australian National University in 2021 and is now a Postdoctoral Researcher at Uppsala University in Sweden. His research interests sit at the intersection of stellar and exoplanetary astrophysics, particularly the characterisation of low-mass stars and their planets.

\noindent \textbf{Prof Anna Moore} is the Director of the Australian National University Institute for Space and Vice-Chancellor’s Entrepreneurial Professor. She was Director of the Advanced Instrumentation and Technology Centre (AITC) from 2017-2021 and was a member of the federal government space expert reference group to form the Australian Space Agency. She is the Chief Investigator of the Dynamic Red All Sky Monitoring System (DREAMS). Professor Moore’s special interests are in instrumentation, Antarctic astronomy and transient infrared astronomy.

\noindent \textbf{Aurelie Magniez} is a PhD candidate at Durham University. She completed a Master degree in Space instrumentation at Université Toulouse III in 2017 then another in signal processing at Université Rennes 1 in 2019. Her research interest includes instrumentation in astronomy, adaptive optics and superconductor detectors.

 \noindent \textbf{A/Prof Luca Casagrande} works in stellar and Galactic astronomy, focusing on the determination of stellar parameters and the use of stellar populations to understand the formation and evolution of the Milky Way. Among his results, he has solved the decade long problem of firmly setting the zero-point of the stellar effective temperature scale, and provided the first observational evidence that stellar radial migration is responsible for broadening the metallicity distribution of stars as a function of age.

\noindent \textbf{Dr Maru{\v s}a {\v Zerjal}} studies young stars near the Sun and investigates their membership in open clusters and associations based on their motions in space. She is using spectroscopic youth indicators and stellar kinematics to constrain their age.

\noindent \textbf{Prof Michael Ireland} is a Professor of Astrophysics and Instrumentation Science at the Australian National University. He obtained his PhD from the University of Sydney in 2006, and has since held positions at the California Insititute of Technology, the Australian Astronomical Observatory, Macquarie University and the University of Sydney. His research focuses on stellar astrophysics, exoplanet formation, the search for life on other worlds and technologies needed to support these endeavours.

\noindent \textbf{Prof Michael Bessell}
is a Professor Emeritus at ANU specializes in the study and evolution of stars, through spectroscopy, photometry and spectrophotometry. A leading expert of photometric systems and their calibration, and the temperature scale of stars, especially stars cooler than the sun.

\noindent \textbf{Nicholas Herald} is an opto-mechanical engineer with 11 years experience, working for Research 
School of Astrophysics and Astronomy at the Australian National University.

\noindent \textbf{Shanae King} is an electronics and embedded systems engineer with experience in the Australian space and defence industry. In 2014, she obtained a bachelor’s degree in Systems Engineering majoring in Mechatronics, and Science majoring in Computer Science, from the Australian National University. Her interests lie in the application of Field Programmable Gate Array (FPGA) and related technologies in space and satellite systems. 

\noindent \textbf{Dr Thomas Nordlander} is a Stromlo Fellow at the Australian National University (ANU). He previously worked at Uppsala University in Sweden, where he obtained his PhD in 2017. His research is focused on stellar spectroscopy and in particular the chemical composition of the oldest stars in the galaxy, as well as cool stars, atomic properties for spectroscopy, and advanced 3D radiative transfer. 

\noindent Biographies of the other authors are not available.


\end{spacing}
\end{document}